%% file: silent-step-spectroscopy_express.tex
\documentclass[submission,copyright,creativecommons]{eptcs}
\providecommand{\event}{EXPRESS/SOS 2024} 
\usepackage{mathtools}
\usepackage{stmaryrd}
\usepackage{extarrows}
\usepackage{centernot}
\usepackage{mathpartir}
  
\usepackage{tabto}
\usepackage{enumitem}
\usepackage{tikz}
\usetikzlibrary{positioning}
\usetikzlibrary{arrows,automata}
\usetikzlibrary{decorations}
\usetikzlibrary{arrows.meta}
\usetikzlibrary{shapes.geometric}
\usetikzlibrary{shapes.multipart}
\usetikzlibrary{shapes.misc}
\usetikzlibrary{shapes}
\usetikzlibrary{fadings}
\usetikzlibrary{fit}
\usepackage{pgf}
\usepackage{pgfplots}
\pgfplotsset{compat=1.18}
\usepackage{relsize} 
\usepackage{adjustbox}
\usepackage{amssymb}
\usepackage{amsthm}
\newtheorem{theorem}{Theorem}[section]

\theoremstyle{definition}
\newtheorem{definition}{Definition}[section]
\newtheorem{example}{Example}[section]
\theoremstyle{remark}
\newtheorem{remark}{Remark}[section]
\usepackage{color}
\usepackage[createShortEnv, conf={end, restate}]{proof-at-the-end}
\usepackage{orcidlink}
\usepackage{ifthen}
\newboolean{arxivversion}
\setboolean{arxivversion}{false}

\makeatletter
\AtBeginDocument{%
  \def\doi#1{\url{https://doi.org/#1}}}
\makeatother

\ifthenelse{\boolean{arxivversion}}{%
  \title{Linear-Time--Branching-Time Spectroscopy \protect\\ Accounting for Silent Steps%
  \footnote{This report provides the proofs for the paper “One Energy Game for the Spectrum between Branching~Bisimilarity and Weak Trace Semantics”, to appear in the proceeding of EXPRESS/SOS~2024.}}
  \def\titlerunning{Linear-Time--Branching-Time Spectroscopy Accounting for Silent Steps}
}{
  \title{One Energy Game for the Spectrum between Branching~Bisimilarity and Weak Trace Semantics}
  \def\titlerunning{One Energy Game for the Weak Spectrum}
}

\author{Benjamin Bisping\orcidlink{0000-0002-0637-0171}
\institute{Technische Universität Berlin, Germany\\
\url{https://bbisping.de}
\email{benjamin.bisping@tu-berlin.de}}
\and
David~N.~Jansen\orcidlink{0000-0002-6636-3301}
\institute{Key~Laboratory~of~System~Software and \\ State~Key~Laboratory~of~Computer~Science, Institute~of~Software, \\
Chinese~Academy~of~Sciences, Beijing, China
\email{dnjansen@ios.ac.cn}}
}

\def\authorrunning{B. Bisping \& D.N. Jansen}

\begin{document}
\input{macros}

\maketitle

\begin{abstract}
  We provide the first generalized game characterization of van Glabbeek's linear-time--branching-time spectrum with silent steps.
  Thereby, \emph{one} multi-dimensional energy game can be used to characterize and decide a wide array of weak behavioral equivalences between stability-respecting branching bisimilarity and weak trace equivalence in one go.
  To establish correctness, we relate at\-tacker-win\-ning energy budgets and distinguishing sublanguages of Hen\-nes\-sy--Mil\-ner logic that we characterize by eight dimensions of formula expressiveness.
\end{abstract}

\section{Introduction: Mechanizing the Spectrum}

Picking the right notion of behavioral equivalence for a particular use case can be hard.%
\footnote{Some accounts of researchers who struggled to pick fitting equivalence for verification and encoding challenges: \cite{concur2022tot2021interview,bell2013certifiably,hm2021ePassportBisim}.}
Theoretically, van Glabbeek's ``linear-time--branching-time spectrum''~\cite{glabbeek1990ltbt1,glabbeek1993ltbt,glabbeek2001ltbtsiReport} brings order to the zoo of equivalences by casting them as a hierarchy of modal logics.
But practically, it is difficult to navigate in particular the second part~\cite{glabbeek1993ltbt}, which considers so-called \emph{weak equivalences} that abstract from ``internal'' behavior, expressed by ``silent'' $\tau$-steps.
Abstracting internal behavior is crucial to model communication happening without participation of the observer and refinements,
that is, for virtually every application.

In this paper, we show how to \emph{operationalize the silent-step linear-time--branching-time spectrum} of~\cite{glabbeek1993ltbt}.
We enable researchers to provide a set of processes that ought to be equated (or distinguished) for their scenario and to learn ``where'' in the spectrum this set of (in-)equivalences holds.
In prior work on the strong spectrum~\cite{glabbeek1990ltbt1} (without silent steps), we dubbed this process \emph{linear-time--branching-time spectroscopy}~\cite{bjn2022decidingAllBehavioralEqs}.
Implicitly, we obtain decision procedures (and games) for each individual notion of equivalence as a by-product.

\begin{figure}[t]
  \begin{adjustbox}{scale=1.0, center}
    \begin{tikzpicture}[->,auto,node distance=3.6cm,
      algstep/.style={minimum width=2.5cm, draw=gray, rectangle,align=center,rounded corners}]
      \node[algstep] (Preord) {$p$ is preordered to $q$\\w.r.t.\@ notion of\\equivalence $N$};
      \node[right=.1cm of Preord] (Piff) {$:\!\iff$};
      \node[align=center, below=.8cm of Piff] (RvG) {van Glabbeek's\\spectrum approach~\cite{glabbeek1993ltbt}};
      \node[algstep, right=.1cm of Piff] (NoFormula) {no formula below $e_N$\\distinguishes $p$ from $q$\\
      (\autoref{sec:background})};
      \node[right=.1cm of NoFormula] (Giff) {$\iff$};
      \node[align=center, below=.8cm of Giff] (SpectroscopyApproach) {Bisping's spectroscopy\\approach~\cite{bisping2023equivalenceEnergyGames} (\autoref{sec:correctness})};
      \node[algstep, right=.1cm of Giff] (DefenderWins) {defender wins spectroscopy game\\ $\gameSpectroscopy$ from $\attackerPos[]{p,\{q\}}$ with energy $e_N$\\
      (\autoref{sec:enrgy-game})};
      \path
        (RvG) edge node {} (Piff)
        (SpectroscopyApproach) edge node {} (Giff);
    \end{tikzpicture}
  \end{adjustbox}
  \caption{How the paper combines the weak spectrum~\cite{glabbeek1993ltbt} and the spectroscopy approach~\cite{bisping2023equivalenceEnergyGames}.}
  \label{fig:overview}
\end{figure}
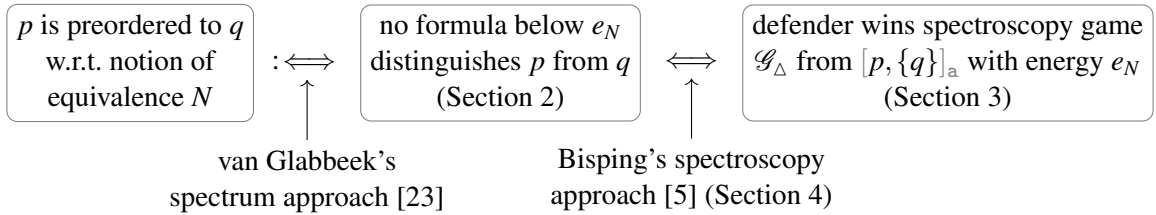

As outlined in \autoref{fig:overview}, we apply our recent approach~\cite{bjn2022decidingAllBehavioralEqs,bisping2023equivalenceEnergyGames} to use a \emph{generalized bisimulation game}
with moves corresponding to sets of conceivable distinguishing formulas.
The background is that \emph{formulas can be partially ordered by the amount of Hennessy--Milner logic expressiveness} they use in a way that aligns with the spectrum.
The game can then be understood as a \emph{multi-weighted energy game}~\cite{bisping2023equivalenceEnergyGames,fjls2011energyGamesMulti,kh2022energyGamesResourceBounded} where moves use up attacker's resources to distinguish processes.
So, defender-won energy levels reveal non-dist\-in\-gui\-shing subsets of Hennessy--Milner logic (HML) and thus sets of maintained equivalences.

Applying the above approach to the weak spectrum faces many obstacles:
The modal logics of the weak spectrum in~\cite{glabbeek1993ltbt} are quite intricate and are not closed under HML-subterms.
Also, van Glabbeek~\cite{glabbeek1993ltbt} does not account for unstable linear-time equivalences, but other publications like Gazda et al.~\cite{gfm2020congruenceOperator} use these.
On the game side, existing weak bisimulation games by De Frutos Escrig et al.~\cite{ekw2017gamesBisimAbstraction} and Bisping et al.~\cite{bnp2020coupledsim30} lack moves for many observations that are relevant for weaker notions in the spectrum.
This paper shows how all this can still be brought together.

\paragraph*{Contributions.}
At its core, this paper extends the spectroscopy energy game of~\cite{bisping2023equivalenceEnergyGames} by modalities needed to cover the weak equivalence spectrum of~\cite{glabbeek1993ltbt}, namely, delayed observations, stable conjunctions, and branching conjunctions.
More precisely:

\begin{itemize}
  \item In \autoref{sec:background}, we capture a big chunk of the \emph{linear-time--branching-time spectrum with silent steps by measuring expressive powers} used in an HML-subset, which we prove to correspond to stability-respecting branching bisimilarity.
  \item In \autoref{sec:enrgy-game}, we introduce the first generalized game characterization of the silent-step equivalence spectrum. For this, we adapt the \emph{spectroscopy energy game} of~\cite{bisping2023equivalenceEnergyGames} to account for distinctions in terms of delayed observations ($\hmlEps\hmlObs{a}\ldots$), stable conjunctions ($\hmlEps\hmlAndS\{\hmlNeg\hmlObs{\tau}\hmlTrue, \ldots\}$), and branching conjunctions ($\hmlEps\hmlAndS\{\hmlObs{a}\ldots, \hmlEps\ldots\}$).
  \item \autoref{sec:correctness} proves that \emph{winning energy levels and equivalences coincide} by closely relating distinguishing formulas and ways the attacker may win the energy game.
  The proofs have been Isabelle/HOL-formalized.
  \item \autoref{sec:algo-refinements} lays out how to use the game to \emph{decide all equivalences at once} in exponential time using our prototype tool for everyday research.
\end{itemize}

\section{Distinctions and Equivalences in Systems with Silent Steps}
\label{sec:background}

This paper follows the paradigm that \emph{equivalence is the absence of possibilities to distinguish}.
Equivalently, one could speak about apartness, i.e.\@ the view that non-equivalence is based on evidence of difference~\cite{geuvers2022}.
We begin by introducing
distinguishing Hennessy--Milner logic formulas (\autoref{subsec:transition-systems}),
and a quantitative characterization of weak equivalences in terms of distinctive capabilities (\autoref{subsec:notions-equivalence}).

\begin{figure}
  \centering
  \begin{adjustbox}{scale=1.0, center}
    \begin{tikzpicture}[->,auto,node distance=2cm, rel/.style={dashed,font=\it, blue}, ext/.style={line width=1pt}]

      \node (Pe){$\ccsIdentifier{P_e}$};
      \node (Ae) [below left of=Pe, node distance=1.5cm] {$\ccsIdentifier{A_e}$};
      \node (Be) [below right of=Pe, node distance = 1.5cm] {$\ccsIdentifier{B_e}$};
      \node (Se) [below right of=Ae, node distance = 1.5cm] {$\circ$};
      \path
      (Pe) edge [bend right=10, swap, ext] node {$\action{op}$} (Ae)
      (Pe) edge [bend left=10, ext] node {$\action{op}$} (Be)
      (Ae) edge [bend right=10, swap, ext] node {$\action{a}\vphantom{\action b}$} (Se)
           edge [loop left, ext] node {$\action{idle}$} ()
      (Be) edge [bend left=10, ext] node {$\action{b}$} (Se)
           edge [loop right, ext] node {$\action{idle}$} ()
      ;

      \node (Pl) [right of=Pe, node distance=6cm] {$\ccsIdentifier{P_\ell}$};
      \node (Al) [below left of=Pl, node distance=1.5cm] {$\ccsIdentifier{A_\ell}$};
      \node (Bl) [below right of=Pl, node distance = 1.5cm] {$\ccsIdentifier{B_\ell}$};
      \node (Sl) [below right of=Al, node distance = 1.5cm] {$\circ$};
      \path
      (Pl) edge [bend right=10, swap, ext] node {$\action{op}$} (Al)
      (Pl) edge [bend left=10, ext] node {$\action{op}$} (Bl)
      (Al) edge [bend right=10, swap, ext] node {$\action{a}\vphantom{\action b}$} (Sl)
           edge [bend right=12, swap, ext] node {$\action{idle}$} (Bl)
           edge [loop left, ext] node {$\action{idle}$} ()
      (Bl) edge [bend left=10, ext] node {$\action{b}$} (Sl)
           edge [bend right=12, swap, ext] node {$\action{idle}$} (Al)
           edge [loop right, ext] node {$\action{idle}$} ()
      ;

      \node (Pte) [below right = .5cm and .9cm of Be]{$\ccsIdentifier{P^\tau_e}$};
      \node (Ate) [below left of=Pte, node distance=1.5cm] {$\ccsIdentifier{A^\tau_e}$};
      \node (Bte) [below right of=Pte, node distance = 1.5cm] {$\ccsIdentifier{B^\tau_e}$};
      \node (Ste) [below right of=Ate, node distance = 1.5cm] {$\circ$};
      \path
      (Pte) edge [bend right=10, swap, ext] node {$\action{op}$} (Ate)
      (Pte) edge [bend left=10, ext] node {$\action{op}$} (Bte)
      (Ate) edge [bend right=10, swap, ext] node {$\action{a} \vphantom{\action b}$} (Ste)
            edge [loop left] node {$\tau$} ()
      (Bte) edge [bend left=10, ext] node {$\action{b}$} (Ste)
            edge [loop right] node {$\tau$} ()
      ;

      \node (Ptl) [right of=Pte, node distance=6cm] {$\ccsIdentifier{P^\tau_\ell}$};
      \node (Atl) [below left of=Ptl, node distance=1.5cm] {$\ccsIdentifier{A^\tau_\ell}$};
      \node (Btl) [below right of=Ptl, node distance = 1.5cm] {$\ccsIdentifier{B^\tau_\ell}$};
      \node (Stl) [below right of=Atl, node distance = 1.5cm] {$\circ$};
      \path
      (Ptl) edge [bend right=10, swap, ext] node {$\action{op}$} (Atl)
      (Ptl) edge [bend left=10, ext] node {$\action{op}$} (Btl)
      (Atl) edge [bend right=10, ext, swap] node {$\action{a} \vphantom{\action b}$} (Stl)
           edge [bend right=10, swap] node {$\tau$} (Btl)
           edge [loop left] node {$\tau$} ()
      (Btl) edge [bend left=10, ext] node {$\action{b}$} (Stl)
           edge [bend right=10, swap] node {$\tau$} (Atl)
           edge [loop right] node {$\tau$} ()
      ;

    \end{tikzpicture}
  \end{adjustbox}
  \caption{A pair of processes $\ccsIdentifier{P_e}$ and $\ccsIdentifier{P_\ell}$ together with versions $\ccsIdentifier{P^\tau_e}$ and $\ccsIdentifier{P^\tau_\ell}$ of the two where $\action{idle}$ has been abstracted into internal $\tau$-behavior.}
  \label{fig:abstracted-processes}
\end{figure}

\subsection{Transition Systems and Hennessy--Milner Logic}
\label{subsec:transition-systems}

\begin{definition}[Labeled transition system with silent steps]
  \label{def:transition-system}
  A \emph{labeled transition system} is a tuple $\system=(\proc,\actions,\step{})$ where $\proc$ is the set of \emph{processes,} $\actions$ is the set of \emph{actions,} and ${\step{}}\subseteq \proc\times\actions\times \proc$ is the \emph{transition relation}.

  $\tau \in \actions$ labels \emph{silent steps} and $\stepWeak$ is notation for the reflexive transitive closure of \emph{internal activity} $\step{\tau}^*$.
  The name $\varepsilon \notin \actions$ is reserved and indicates no (visible) action.
  A process $p$ is called \emph{stable} if \mbox{$p \centernot{\step{\tau}}$}.
  We write $p \step{\hmlOpt{\alpha}} p'$ if $p \step{\alpha} p'$, or if $\alpha = \tau$ and $p = p'$.

  We implicitly lift the relations to sets of processes $P \step{\alpha} P'$ (with $P, P' \subseteq \proc$, $\alpha \in \actions$), which is defined to be true if $P' = \{ p' \in \proc \mid \exists p \in P \ldotp p \step{\alpha} p'\}$.
\end{definition}

\begin{example}
  \label{exa:abstracted-processes}
  \autoref{fig:abstracted-processes} presents transition systems of four processes:
  $\ccsIdentifier{P_e}$ makes a nondeterministic choice $\action{op}$ between $\action{a}$ and $\action{b}$, performing arbitrarily many $\action{idle}$-actions in between.
  $\ccsIdentifier{P_\ell}$ does the same but can change the choice while idling.
  $\ccsIdentifier{P^\tau_e}$ and $\ccsIdentifier{P^\tau_\ell}$ are variants of the two obtained by abstracting $\action{idle}$ into $\tau$-actions.

  The example is helpful to test whether a process equivalence can be a congruence for abstraction.
  Any congruence for abstraction $\sim$ would need to have the property that $\ccsIdentifier{P_e} \sim \ccsIdentifier{P_\ell}$ implies $\ccsIdentifier{P^\tau_e} \sim \ccsIdentifier{P^\tau_\ell}$.
  So, if we just had a quick way of testing for all weak behavioral equivalences at once, we could quickly narrow down which equivalences work for this example.
  Using this paper's spectroscopy algorithm, we can achieve this.
\end{example}

\definecolor{etaColor}{cmyk}{0.8, 0, 0, 0.3}
\definecolor{stabilityColor}{cmyk}{0, 0.7, 0.7, 0.4}

\noindent
Bisimilarity and other notions of equivalence can conveniently by defined in terms of Hennessy--Milner logic.
We direct our attention to variants that allow for silent behavior to happen before visible actions are observed.
We thus focus on the following variant,
where the \textcolor{stabilityColor}{brick-red} part represents \emph{stable conjunctions}
and the \textcolor{etaColor}{steel-blue} part \emph{branching conjunctions}:

\begin{definition}[Branching Hennessy--Milner logic]
  \label{def:hml-branching}
  We define \emph{stability-respecting branching} Hennessy--Milner modal logic,
  $\hmlB$, over an alphabet of actions $\actions$ by the following context-free grammar starting with~$\varphi$:
  \begin{displaymath}
    \arraycolsep=.4pt\def\arraystretch{1.5}
    \begin{array}{c@{\;\displaystyle}llr}
      \varphi {} ::= {} & \hmlEps\chi & & \text{“delayed observation”} \\*
          | \quad & \hmlAndS \{\psi, \psi, ...\} & & \text{“immediate conjunction”} \\
      \chi {} ::= {} & \hmlObs{a}\varphi & \quad\text{with } a \in \actions \setminus \{\tau\} & \text{“observation”} \\*
        | \quad & \hmlAndS \{\psi, \psi, ...\} & & \text{“standard conjunction”} \\*
        | \quad & \textcolor{stabilityColor}{\hmlAndS \{\neg\hmlObs{\tau}\hmlTrue, \psi, \psi, ...\}} & &
          \textcolor{stabilityColor}{\text{“stable conjunction”}} \\*
        | \quad & \textcolor{etaColor}{\hmlAndS \{\hmlOpt{\alpha}\varphi, \psi, \psi, ...\}} &
        \quad\textcolor{etaColor}{\text{with } \alpha \in \actions} & \textcolor{etaColor}{\text{“branching conjunction”}} \\
      \psi {} ::= {} & \hmlNeg\hmlEps\chi \mid \hmlEps\chi & & \text{“negative / positive conjuncts”}
    \end{array}
  \end{displaymath}
  Its semantics $\;\smash{\hmlSemantics{\;\cdot\;}{\system}{}} \colon \hmlB \to \powerSet{\proc}\!$,
  where a formula ``is true,''
  over a transition system $\system=(\proc,\actions,\step{})$
  is defined in mutual recursion with helper functions $\hmlSemantics{\;\cdot\;}{}{\varepsilon}$ for subformulas in the ``delayed'' context ($\chi$-productions) and $\hmlSemantics{\;\cdot\;}{}{\wedge}$ for conjuncts ($\psi$-productions):
  \begin{align*}
    \hmlSemantics{\hmlEps\chi}{\system}{} \defEq &\;
      \{p \in \proc \mid
        \exists p' \in \hmlSemantics{\chi}{\system}{\varepsilon} \ldotp p \stepWeak p'\}
    \\[0.25\baselineskip]
    \smash{\hmlSemantics{\hmlAndS \Psi}{\system}{}} \defEq \smash{\hmlSemantics{\hmlAndS \Psi}{\system}{\varepsilon}} \defEq &\;
          \bigcap \{ \hmlSemantics{\psi}{\system}{\wedge} \mid
            \psi \in \Psi \}
    \\[0.25\baselineskip]
    \hmlSemantics{\hmlObs{a}\varphi}{\system}{\varepsilon} \defEq &\;
      \{p \in \proc \mid
        \exists p' \in \hmlSemantics{\varphi}{\system}{} \ldotp p \step{a} p'\}
    \\[0.25\baselineskip]
    \hmlSemantics{\textcolor{stabilityColor}{\hmlNeg\hmlObs{\tau}\hmlTrue}}{\system}{\wedge} \defEq &\;
      \{p \in \proc \mid
        p \centernot{\step{\tau}} \}
    \\[0.25\baselineskip]
    \hmlSemantics{\textcolor{etaColor}{\hmlOpt{\alpha}\varphi}}{\system}{\wedge} \defEq &\;
      \{p \in \proc \mid
        \exists p' \in \hmlSemantics{\varphi}{\system}{} \ldotp p \step{\hmlOpt{\alpha}} p'\}
    \\[0.25\baselineskip]
    \hmlSemantics{\hmlNeg\hmlEps\chi}{\system}{\wedge} \defEq &\;
      \proc \setminus \hmlSemantics{\hmlEps\chi}{\system}{}
    \\[0.25\baselineskip]
    \hmlSemantics{\hmlEps\chi}{\system}{\wedge} \defEq &\;
      \hmlSemantics{\hmlEps\chi}{\system}{}
  \end{align*}
\end{definition}

\noindent
$\hmlAndS \{\psi, \psi, ...\}$ in the grammar stands for conjunction with arbitrary branching.
We write $\hmlTrue$ for the empty conjunction $\hmlAndS \varnothing$.

\begin{definition}[Distinguishing formulas and preordering languages]
  \label{def:distinguishing-formula}
  A formula $\varphi \in \hmlB$ is said to \emph{distinguish} a process $p$ from $q$ iff $p \in \hmlSemantics{\varphi}{\smash{\system}}{}$ and $q \notin \smash{\hmlSemantics{\varphi}{\smash{\system}}{}}$\!.
  The formula is said to \emph{distinguish} a process $p$ from a set of processes $Q$ iff it is  true for $p$ and false for every $q \in Q$.
  
  A sublogic, $\obs{\mathit N} \subseteq \hmlB$, corresponding to a notion of observability $N$, \emph{distinguishes} two processes, $p \not\bPreord{\mathit N} q$, if there is $\varphi \in \obs{\mathit N}$ with $p \in \hmlSemantics{\varphi}{\smash{\system}}{}$ and $q \notin \smash{\hmlSemantics{\varphi}{\smash{\system}}{}}$. Otherwise $N$ \emph{preorders} them, $p \bPreord{\mathit N} q$. 
  If processes are mutually $N$-preordered, $p \bPreord{\mathit N} q$ and $q \bPreord{\mathit N} p$, then they are considered $N\!$-\emph{equivalent}, $p \bEquiv{\mathit N} q$.
\end{definition}

\begin{example}
  \label{exa:distinguishing-formula}
  In \autoref{exa:abstracted-processes}, $\varphi_\tau \defEq \hmlEps\hmlObsI{op}\hmlEps\hmlAndS\{\hmlNeg\hmlEps\hmlObsI{b}\hmlTrue \}$ distinguishes $\ccsIdentifier{P^\tau_e}$ from $\ccsIdentifier{P^\tau_\ell}$.
  $\varphi_\tau$ states that a weak $\action{op}$-step may happen such that, afterwards, $\action{b}$ is not $\tau$-reachable.
  This is true of $\ccsIdentifier{P^\tau_e}$ because of the $\ccsIdentifier{A^\tau_e}$-state, but not of $\ccsIdentifier{P^\tau_\ell}$.
\end{example}

\begin{remark}
  \autoref{def:hml-branching} is constructed to fit the distinctive powers we need from HML to characterize varying notions of the weak spectrum by controlling which productions are used.
  Subformulas in the grammar usually start with $\hmlEps\ldots$,
  effectively hiding silent steps.
  Formulas with fewer $\hmlEps$-positions bring in additional distinctive power.
  We will use immediate conjunctions to distinguish non-delay-bisimilar processes, and branching conjunctions (that contain one positive conjunct without leading $\hmlEps$) to distinguish non-$\eta$-(bi)similar processes.
  Allowing the observation of stabilization, $\textcolor{stabilityColor}{\hmlNeg\hmlObs{\tau}\hmlTrue}$, increases distinctive power; requiring stabilization for conjunct observations decreases it.
\end{remark}

\noindent
The name already alludes to $\hmlB$ as a whole characterizing stability-re\-spect\-ing branching bisimilarity.
Let us quickly recall the operational definition for branching bisimilarity (for instance from~\cite{FokkinkGL19DivCong3}):

\begin{definition}[Branching bisimilarity, operationally]
  \label{def:branching-bisim-operationally}
  A symmetric relation $\rel{R}$ is a \emph{branching bisimulation}
  if, for all $(p,q) \in \rel R$, a step $p \step{\alpha} p'$ implies (1)~$\alpha = \tau$ and $(p', q) \in \rel R$, or (2)~$q \stepWeak q' \step{\alpha} q''$ for some $q', q''$ with $(p, q') \in \rel R$ and $(p', q'') \in \rel R$.

  If moreover every $(p,q) \in \rel R$ with $p \nostep{\tau}$ implies that there is some $q'$ with $q \stepWeak q' \nostep{\tau}$ and $(p, q') \in \rel R$, the relation is \emph{stability-re\-spect\-ing}.
  
  If there is a stability-respecting branching bisimulation $\rel R_{BB^{sr}}$ with $(p_0, q_0) \in \rel R_{BB^{sr}}$, then $p_0$ and $q_0$ are stability-respecting branching bisimilar.
\end{definition}

\noindent
The power of \autoref{def:hml-branching} to distinguish matches exactly the power of \autoref{def:branching-bisim-operationally} to equate:

\begin{lemmaE}[][see full proof, restate]
  \label{prop:srbb-characterization}
  $\hmlB$ characterizes stability-respecting branching bisimilarity.
  \ifthenelse{\boolean{arxivversion}}{
    (i.e.\@ $p$ is not stability-respecting branching bisimilar to $q$ iff there exists a formula in $\hmlBA$ that distinguish $p$ from $q$ or $q$ from $p$, respectively.)
  }{}
\end{lemmaE}
\begin{proof}
We use the standard approach for Hennessy--Milner theorems:
We prove that $\rel{R}_{srbb} \defEq \{ (p,q) \mid \forall \varphi \in \hmlB \ldotp p \in \hmlSemantics{\varphi}{}{} \longrightarrow q \in \hmlSemantics{\varphi}{}{}\}$ is a stability-respecting branching bisimulation by definition,
and that any formula $\varphi \in \hmlB$ is equally true for stability-respecting branching bisimilar states by induction on the structure of $\varphi$.
\ifthenelse{\boolean{arxivversion}}{
\begin{proofE}
We prove that there is no formula $\varphi \in \hmlB$ distinguishing $p_0$ from $q_0$ if and only if there is a stability-respecting branching bisimulation $\rel R$ by \autoref{def:branching-bisim-operationally} with $(p_0,q_0) \in \rel R$.

Assume no $\varphi \in \hmlB$ distinguishes $p_0$ from $q_0$.
Consider $\rel{R}_{srbb} \defEq \{ (p,q) \mid \forall \varphi \in \hmlB \ldotp p \in \hmlSemantics{\varphi}{}{} \longrightarrow q \in \hmlSemantics{\varphi}{}{}\}$.
Clearly, $(p_0, q_0) \in \rel{R}_{srbb}$.
We will show $\rel{R}_{srbb}$ to be a stability-respecting branching bisimulation.
\begin{itemize}
  \item Symmetry of $\rel{R}_{srbb}$ (by contradiction):
    Assume $\rel{R}_{srbb}$ were not symmetric.
    Then there were $(p,q) \in \rel{R}_{srbb}$ with $(q,p) \notin \rel{R}_{srbb}$.
    The latter means there is $\varphi \in \hmlB$ with $q \in \hmlSemantics{\varphi}{}{}$ and $p \notin \hmlSemantics{\varphi}{}{}$.
    Consider the cases of $\varphi$:
    \begin{itemize}
      \item $\varphi = \hmlEps\chi$.
        Then $\hmlAndS \{ \hmlNeg \hmlEps \chi \} \in \hmlB$ distinguishes $p$ from $q$, contradicting $(p,q) \in \rel{R}_{srbb}$.
      \item $\varphi = \hmlAndS \Psi$.
        That means that there is some $\psi \in \Psi$ such that $p \notin \hmlSemantics{\psi}{}{}$,
        while $q \in \hmlSemantics{\hmlAndS \Psi}{}{}$ implies $q \in \hmlSemantics{\psi}{}{}$.
        If $\psi = \hmlNeg\hmlEps\chi$, then $\hmlEps\chi \in \hmlB$ distinguishes $p$ from $q$.
        But if $\psi = \hmlEps\chi$, then $\hmlAndS \{ \hmlNeg\hmlEps\chi \} \in \hmlB$ distinguishes $p$ from $q$.
        In both cases we have a contradiction to $(p,q) \in \rel{R}_{srbb}$.
    \end{itemize}
  \item Respect of stability (by contradiction):
    Assume there were $(p,q) \in \rel{R}_{srbb}$ with $p \nostep{\tau}$ but all $q'$ with $q \stepWeak q'$ having $q' \step{\tau} q''$ or $(p, q') \notin \rel{R}_{srbb}$.
    For each $q' \nostep{\tau}$, there must be a distinguishing formula $\varphi_{q'}$.
    The others can be distinguished from $p$ by the fragment $\hmlNeg\hmlObs{\tau}\hmlTrue$.
    For all $\varphi_{q'}$ that are conjunctions, let $\psi_{q'}$ denote some conjunct that distinguishes $p$ from $q'$ (some must exist);
    and let $\psi_{q'} \defEq \varphi_{q'}$ for the others.
    Then, $\hmlEps\hmlAndS\{\hmlNeg\hmlObs{\tau}\hmlTrue\} \cup \{ \psi_{q'} \mid q \stepWeak q' \nostep{\tau} \} \in \hmlB$ distinguishes $p$ from $q$, contradicting $(p,q) \in \rel{R}_{srbb}$.
  \item Branching simulation on $\step{a}$ (by contradiction):
    Assume $p \step{a} p'$, $(p,q) \in \rel{R}_{srbb}$, but for all $q'$ and $q''$ with $q \stepWeak q' \step{a} q''$, $(p,q') \notin \rel{R}_{srbb}$ or $(p',q'') \notin \rel{R}_{srbb}$.
    Let us refer to the respective distinguishing formulas as $\varphi_\varepsilon(q')$ (with $q \stepWeak q'$) and define $Q_a$ as the set of $q'$ that cannot be distinguished from $p$, but where $\varphi_a(q',q'')$ distinguishes $p'$ from any $q''$ with $q' \step{a} q''$.
    As in the previous case, let $\psi_\epsilon(q')$ and $\psi_a(q',q'')$ refer to specific distinguishing conjuncts in $\varphi_\epsilon(q')$ and $\varphi_a(q',q'')$.
    Consider
      $\varphi_\eta = \hmlEps \hmlAndS \{ \hmlObs{a} \hmlAndS \{ \psi_a(q', q'') \mid  q' \in Q_a, q' \step{a} q'' \} \} \cup
      \{ \psi_\varepsilon(q') \mid q \stepWeak q', q' \notin Q_a \}$,
    $\varphi_\eta$ combines all the distinctions for the $q$-derivatives in a strengthened formula, which must still be true for $p$ because we know of the $p \step{a} p'$ transition.
    For the $q$-side, this is sound because whatever distinctions render the follow-up formulas false for $q'$ / $q''$, must be included in $\varphi_\eta$.
    Thus $\varphi_\eta$ distinguishes $p$ from $q$, contradicting $(p,q) \in \rel{R}_{srbb}$.
  \item Branching simulation on $\step{\tau}$ (by contradiction):
    Assume $p \step{\tau} p'$ and $(p,q) \in \rel{R}_{srbb}$.
    If $(p',q) \in \rel{R}_{srbb}$,
    then we are finished.
    Otherwise, we can derive a contradiction
    by constructing a similar branching conjunction formula as in the previous case.
    The difference is that $\varphi_\tau(q',q'')$ should not only distinguish $p'$ from $q''$ but also from $q'$;
    this is possible because we can use $(p',q) \notin \rel{R}_{srbb}$.
\end{itemize}

\noindent
Assume $p_0$ and $q_0$ are stability-respecting branching bisimilar.
This means $(p_0, q_0) \in \rel R_{BB^{sr}}$, the greatest stability-respecting branching bisimulation.
Assume moreover that $p_0 \in \hmlSemantics{\varphi}{}{}$.
We will show that this implies $q_0 \in \hmlSemantics{\varphi}{}{}$ by induction over the structure of $\varphi$ (and inner $\psi$) with arbitrary $p_0$ and $q_0$.
\begin{itemize}
  \item Case $\varphi = \hmlEps\chi$.
    $p \in \hmlSemantics{\hmlEps\chi}{}{}$ implies there are $p \step{\tau}^n p'$ such that $p' \in \hmlSemantics{\chi}{}{}$.
    To prove $q \in \hmlSemantics{\hmlEps\chi}{}{}$, we will establish that there is $q' \in \hmlSemantics{\chi}{}{}$ with $q \stepWeak q'$ by considering the cases for $\chi$:
    \begin{itemize}
      \item Case $\chi = \hmlObs{a}\varphi'$.
        This implies there is $p''$ with $p' \step{a} p''$ and $p'' \in \hmlSemantics{\varphi'}{}{}$.
        Due to branching bisimulation, there are $q', q''$ with $q \stepWeak \stepWeak q' \step{a} q''$, $(p', q') \in \rel R_{BB^{sr}}$, and $(p'', q'') \in \rel R_{BB^{sr}}$.
        With the induction hypothesis, this implies $q'' \in \hmlSemantics{\varphi'}{}{}$, and due to the HML semantics, $q' \in \hmlSemantics{\hmlObs{a}\varphi'}{}{}$.
      \item Case $\chi = \hmlAndS \Psi$.
        We know that each $\psi \in \Psi$ must be true for $p'$.
        If we choose an appropriate $q'$ with $q \stepWeak q'$ and $(p', q') \in \rel R_{BB^{sr}}$,
        the induction hypothesis implies each $\psi$ of the form $\hmlNeg\hmlEps\chi'$ or $\hmlEps\chi'$ to be true for $q'$ as well.

        If there is $\psi = \hmlNeg\hmlObs{\tau}\hmlTrue \in \Psi$, its truth ensures $p' \nostep{\tau}$,
        we choose $q'$ to be one where $q \stepWeak q' \nostep{\tau}$ and $(p', q') \in \rel R_{BB^{sr}}$, thanks to $\rel R_{BB^{sr}}$ respecting stability.
        $q'$ satisfies $\hmlNeg\hmlObs{\tau}\hmlTrue$.
        
        If, otherwise, there is $\psi = \hmlOpt{\alpha}\varphi' \in \Psi$ (implying $p'' \in \hmlSemantics{\varphi'}{}{}$ and $p' \step{(\alpha)} p''$),
        we choose $q'$ to be one where $q \stepWeak q' \step{(\alpha)} q''$, $(p',q') \in \rel R_{BB^{sr}}$, and $(p'',q'') \in \rel R_{BB^{sr}}$,
        thanks to $\rel R_{BB^{sr}}$ being a branching simulation.
        By induction hypothesis, $\varphi'$ must be true for $q''$ as well
        and thus  $\hmlOpt{\alpha}\varphi'$ holds for this $q'$.

        If neither of the prior two conjuncts are present, we just take a $q' = q$ if $p \nostep{\tau}$,
        or otherwise some $q'$  with $q \stepWeak q'$ and $(p', q') \in \rel R_{BB^{sr}}$ that is implied by the simulation property of $\rel R_{BB^{sr}}$ on $p \step{\tau}^n p'$.
        
        In every case, we have found a  $q' \in \hmlSemantics{\hmlAndS \Psi}{}{}$.
    \end{itemize}
    
  \item Case $\varphi = \hmlAndS \Psi$.
    We know that each $\psi \in \Psi$ must be true for $p$.
    By induction hypothesis, $(p,q) \in R_{BB^{sr}}$ implies each $\psi$ to be true for $q$ as well.
    Thus $q \in \hmlSemantics{\hmlAndS \Psi}{}{}$.

  \item Case $\psi = \hmlEps\chi$.
    The proof of the first case also addresses this case.
    
  \item Case $\psi = \hmlNeg\hmlEps\chi$.
    Thanks to the proof of the first case and symmetry of $\rel R_{BB^{sr}}$, we know that if $\hmlEps\chi$ were to be true for $q$, it would also need to be true for $p$.
    The case implies that $p \notin \hmlSemantics{\hmlEps\chi}{}{}$.
    By contraposition, $q \notin \hmlSemantics{\hmlEps\chi}{}{}$.
    This proves that $\hmlNeg\hmlEps\chi$, in the context of a conjunction, does agree with~$q$. \qedhere
\end{itemize}
\end{proofE}
}{
  Full proof in report~\cite{bj2023silentStepSpectroscopyArxiv}.
}
\end{proof}

\newcommand*{\EqCoords}[1]{\textcolor{gray}{\small $#1$}}
\begin{figure*}[t!]
  \begin{adjustbox}{max width=\textwidth, center}
    \begin{tikzpicture}[auto,node distance=2.5cm,align=center]
      \node (BBs){\color{stabilityColor}stability-respecting \color{etaColor} branching bisim $\eqName{BB^{\color{stabilityColor}sr}}$\\
        \EqCoords{\infty,\infty,\infty,\infty,\infty,\infty,\infty,\infty}};
      \node (BB)[below left of=BBs]{\color{etaColor}branching bisim $\eqName{BB}$\\
        \EqCoords{\infty,\infty,\infty,0,\infty,\infty,\infty,\infty}};
      \node (eB)[xshift=-5mm, below left of=BB]{\color{etaColor}$\eta$-bisim $\eqName{\eta}$\\
        \EqCoords{\infty,\infty,\infty,0,0,\infty,\infty,\infty}};
      \node (DB)[below right of=BB]{delay bisim $\eqName{DB}$\\
        \EqCoords{\infty,0,\infty,0,\infty,\infty,\infty,\infty}};
      \node (B)[below left of=DB]{weak bisim $\eqName{B}$\\
        \EqCoords{\infty,0,\infty,0,0,\infty,\infty,\infty}};
      \node (S2)[below left of=B]{$2$-nested sim $\eqName{2S}$\\
        \EqCoords{\infty,0,\infty,0,0,\infty,\infty,1}};
      \node (C)[below right of=B]{contrasim $\eqName{C}$\\
        \EqCoords{\infty,0,\infty,0,0,0,\infty,\infty}};
      \node (RS)[below left of=S2]{ready sim $\eqName{RS}$\\
        \EqCoords{\infty,0,\infty,0,0,\infty,1,1}};
      \node (R)[below right of=RS]{readiness $\eqName{R}$\\
        \EqCoords{\infty,0,1,0,0,1,1,1}};
      \node (PF)[above right of=R]{~possible futures $\eqName{PF}$\\
        \EqCoords{\infty,0,1,0,0,\infty,\infty,1}};
      \node (IF)[below right of=PF]{impossible futures $\eqName{IF}$\\
        \EqCoords{\infty,0,1,0,0,0,\infty,1}};
      \node (S)[below left of=RS]{sim $\eqName{1S}$\\
        \EqCoords{\infty,0,\infty,0,0,\infty,0,0}};
      \node (eS)[above left of=RS]{\color{etaColor}$\eta$-sim $\eqName{\eta S}$\\
        \EqCoords{\infty,\infty,\infty,0,0,\infty,0,0}};
      \node (F)[below right of=R]{failures $\eqName{F}$\\
        \EqCoords{\infty,0,1,0,0,0,1,1}};
      \node (T)[below left of=F]{traces $\eqName{T}$\\
        \EqCoords{\infty,0,0,0,0,0,0,0}};
      %
      \node (DBs)[xshift=12mm, above right of=DB, node distance=1.75cm]{\color{stabilityColor}s.-r.\ delay bisim $\eqName{DB^{sr}}$\\
        \EqCoords{\infty,0,\infty,\infty,\infty,\infty,\infty,\infty}};
      \node (SB)[below right of=DBs]{\color{stabilityColor}stable bisim $\eqName{SB}$\\
        \EqCoords{\infty,0,0,\infty,0,\infty,\infty,\infty}};
      \node (RSs)[below left of=SB]{\color{stabilityColor}stable ready sim $\eqName{RS^s}$\\
        \EqCoords{\infty,0,0,\infty,0,\infty,1,1}};
      \node (Rs)[below of=RSs, node distance=3cm]{\color{stabilityColor}stable readiness $\eqName{R^s}$\\
        \EqCoords{\infty,0,0,1,0,1,1,1}};
      \node (IFs)[below of=SB, node distance=3cm]{\color{stabilityColor}~~stable imposs.\ fut.\ $\eqName{IF^s}$\\
        \EqCoords{\infty,0,0,1,0,0,\infty,1}};
      \node (Fs)[below right of=Rs]{\color{stabilityColor}stable failures $\eqName{F^s}$\\
        \EqCoords{\infty,0,0,1,0,0,1,1}};
      \path
      (BBs) edge (BB)
      (BB) edge ([xshift=10mm]eB.north)
      (BB) edge (DB)
      ([xshift=10mm]eB.south) edge (B)
      (eB.south) edge (eS)
      (DB) edge (B)
      (B) edge (S2)
      (B) edge (C)
      (C) edge (IF)
      (S2) edge (RS)
      (S2) edge (PF)
      (RS) edge (S)
      (RS) edge (R)
      (PF) edge (R)
      (PF) edge (IF)
      (eS) edge (S)
      (S) edge (T)
      (IF) edge (F)
      (R) edge (F)
      (F) edge (T)
      ;
      \path
      (BBs) edge (DBs)
      ([xshift=-4mm]DBs.south) edge ([xshift=-9mm,yshift=-5mm]DBs.south)
      (DBs) edge (SB)
      (SB) edge (RSs)
      (SB) edge (IFs)
      (RSs) edge (Rs)
      (Rs) edge (Fs)
      (IFs) edge (Fs)
      (Fs) edge ([xshift=30mm]T.east)
      ([xshift=30mm]T.east) edge (T.east)
      ;
    \end{tikzpicture}
  \end{adjustbox}
  \caption[]{Hierarchy of weak behavioral equivalences/preorders, becoming finer towards the top. Each notion $N$ comes with its expressiveness coordinate $e_N$.
  \protect{\ifthenelse{\boolean{arxivversion}}{
    Lines mean implication of equivalence/preordering from bottom to top.
  }{}}
  }
  \label{fig:ltbt-spectrum}
\end{figure*}

\subsection{Price Spectra of Behavioral Equivalences}
\label{subsec:notions-equivalence}

Van Glabbeek~\cite{glabbeek1993ltbt} uses about 20 binary dimensions to characterize 155 ``notions of observability'' (derived from five dimensions of testing scenarios).
These then entail behavioral preorders and equivalences given as modal characterizations.
In this subsection, we recast the \emph{notions of observability as coordinates} in a (more quantitative) 8-dimensional space of HML formula expressiveness.

We will ``price'' formulas of $\hmlB$ by vectors we call \emph{energies}. The pricing allows to conveniently select subsets of $\hmlB$ in terms of coordinates.

\begin{definition}[Energies]
  \label{def:energies}
  We denote as \emph{energies,} $\energies_\infty$, the set $(\nats\cup\{\infty\})^8$.

  We compare energies component-wise: $\vectorComponents[8]{e} \leq \vectorComponents[8]{f}$ iff $e_i \leq f_i$ for each $i$.
  Least upper bounds $\sup$ are defined as usual as component-wise supremum.

  We write $\unit{i}$ for the standard unit vector where the $i$-th component is $1$ and every other component equals $0$. $\zeroVec$ is defined to be the vector $(0,0, \ldots, 0)$. Vector addition and subtraction happen component-wise as usual.
\end{definition}

\noindent
In \autoref{fig:ltbt-spectrum}, we order weak equivalences along dimensions of $\hmlB$-ex\-pres\-si\-ve\-ness in terms of \emph{operator depths} (\ie maximal occurrences of an operator on a path from root to leaf in the abstract syntax tree).
Intuitively, the dimensions are:

\begin{enumerate}[noitemsep]
  \item Modal depth (of observations $\hmlObs{\alpha}$, $\hmlOpt{\alpha}$),
  \item Depth of \textcolor{etaColor}{branching conjunctions} (with one observation conjunct not starting with $\hmlEps$),
  \item Depth of unstable conjunctions (that do not enforce stability by a $\hmlNeg \hmlObs{\tau} \hmlTrue$-conjunct),
  \item Depth of \textcolor{stabilityColor}{stable conjunctions} (that do enforce stability by a $\hmlNeg \hmlObs{\tau} \hmlTrue$-conjunct),
  \item Depth of immediate conjunctions (that are not preceded by $\hmlEps$),
  \item Maximal modal depth of positive conjuncts in conjunctions,
  \item Maximal modal depth of negative conjuncts in conjunctions,
  \item Depth of negations.
\end{enumerate}

\begin{definition}[Formula prices]
  \label{def:formula-prices}
  The \emph{expressiveness price} of a formula $\expr \colon \hmlB \rightarrow \energies_\infty$
  is defined in mutual recursion with helper functions $\expr^\varepsilon$ and $\expr^\wedge$;
  if multiple rules apply to a subformula, pick the first one:
  \begin{align*}
    \expr\left(\hmlTrue\right) \defEq
    \expr^\varepsilon\left(\hmlTrue\right) & \defEq
    \zeroVec
    \\
    \textstyle\expr\left(\hmlEps\chi\right) & \defEq
    \textstyle\expr^\varepsilon\left(\chi\right)
    \\
    \textstyle\expr\left(\hmlAndS\Psi\right) & \defEq
    \unit{5} + \textstyle\expr^\varepsilon\left(\hmlAndS \Psi\right)
    \\
    \expr^\varepsilon\left(\hmlObs{a}\varphi\right) & \defEq
    \unit{1} + \expr\left(\varphi\right)
    \\
    \textstyle\expr^\varepsilon\left(\hmlAndS \Psi\right) & \defEq
    \sup\;\{ \expr^\wedge\left(\psi\right) \mid \psi \in \Psi \} +
    \begin{cases}
      \textcolor{stabilityColor}{\unit{4}} & \text{if } \textcolor{stabilityColor}{\hmlNeg\hmlObs{\tau}\hmlTrue} \in \Psi\\
      \textcolor{etaColor}{\unit{2}} + \unit{3} & \text{if there is } \textcolor{etaColor}{\hmlOpt{\alpha}\varphi} \in \Psi \\
      \unit{3} & \text{otherwise}
    \end{cases}
    \\
    \expr^\wedge\left(\textcolor{stabilityColor}{\hmlNeg\hmlObs{\tau}\hmlTrue}\right) & \defEq
    \zeroVec
    \\
    \expr^\wedge\left(\hmlNeg\varphi\right) & \defEq
    \sup\;\{ \unit{8} + \expr\left(\varphi\right),\quad
      (0,0,0,0,0,0,\left(\expr\left(\varphi\right)\right)_1,0) \}
    \\
    \expr^\wedge\left(\textcolor{etaColor}{\hmlOpt{\alpha}\varphi}\right) & \defEq
    \sup\;\{ \unit{1} + \expr\left(\varphi\right),\quad
      (0,0,0,0,0,1 + \left(\expr\left(\varphi\right)\right)_1,0,0) \}
    \\
    \expr^\wedge\left(\varphi\right) & \defEq
    \sup\;\{ \hphantom{\unit{8} + {}} \expr\left(\varphi\right),\quad
      (0,0,0,0,0,\left(\expr\left(\varphi\right)\right)_1,0,0) \}
  \end{align*}
\end{definition}

\begin{definition}[Linear-time--branching-time equivalences]
  \label{def:language-prices}
  Each notion $N$ named in \refFig{fig:ltbt-spectrum} with coordinate $e_N$ is defined through the language of formulas with prices below, i.e., through $\obs{\mathit N} = \{ \varphi \mid \expr(\varphi) \leq e_N\}$.
\end{definition}
Recalling \autoref{def:distinguishing-formula}, that is, $p \bPreord{\mathit{N}} q$ with respect to notion $N$, iff no $\varphi$ with $\expr(\varphi) \leq e_N$ distinguishes $p$ from $q$.
So, this paper sees notions of preorder / equivalence to be defined through these coordinates and not through other characterizations.

\begin{example}
  \label{exa:distinction-price}
  The formula $\varphi_\tau = \hmlEps\hmlObsI{op}\hmlEps\hmlAndS\{\hmlNeg\hmlEps\hmlObsI{b}\hmlTrue \}$
  in \autoref{exa:distinguishing-formula} has expressiveness price $\expr(\varphi_\tau) = (2,0,1,0,0,0,1,1).$
  The coordinate is below the one of failures $e_{\mathrm{F}} = (\infty,0,1,0,0,0,1,1)$ in \autoref{fig:ltbt-spectrum}.
  Accordingly, $\ccsIdentifier{P^\tau_e}$ is distinguished from $\ccsIdentifier{P^\tau_\ell}$ by failure $\varphi_\tau \in \obs{F}$, that is, $\ccsIdentifier{P^\tau_e} \not\bPreord{F} \ccsIdentifier{P^\tau_\ell}$.
  There neither are strictly-stable nor strictly-positive formulas to distinguish $\ccsIdentifier{P^\tau_e}$ from $\ccsIdentifier{P^\tau_\ell}$. Therefore, stable bisimulation preorder, $\ccsIdentifier{P^\tau_e} \bPreord{SB} \ccsIdentifier{P^\tau_\ell}$, and $\eta$-simulation preorder, $\ccsIdentifier{P^\tau_e} \bPreord{\eta S} \ccsIdentifier{P^\tau_\ell}$, apply.
  (The latter implies the more well-known weak simulation preorder.)
\end{example}

\noindent
For stability-respecting branching bisimilarity, where $\obs{\eqName{BB^{sr}}} = \hmlB$, \autoref{prop:srbb-characterization} establishes that our modal characterization corresponds to the common relational definition.
For some notions, there are superficial differences to other modal characterizations in the literature, which do not change distinctive power.
We give two examples.

\begin{example}[Weak trace equivalence and inclusion]
  The notion of weak trace inclusion (and equivalence) is defined through $e_{\eqName{T}} = (\infty,0,0,0,0,0,0,0)$ and \autoref{def:formula-prices} inducing $\obs{T}$, the language given by the grammar:
  \begin{align*}
    \varphi_\eqName{T} \quad ::= \quad & \quad \hmlEps\hmlObs{a}\varphi_\eqName{T}
                \quad | \quad \hmlEps\hmlTrue 
                \quad | \quad \hmlTrue .
  \end{align*}
  This slightly deviates from languages one would find in other publications. For instance, Gazda et al.~\cite{gfm2020congruenceOperator} do not have the second production. But this production does not increase expressiveness, as $\hmlSemantics{\hmlEps\hmlTrue}{}{} = \hmlSemantics{\hmlTrue}{}{} = \proc$.
\end{example}

\begin{example}[Weak bisimulation equivalence and preorder]
  \label{exa:weak-bisim-modal}
  The logic of weak bisimulation observations $\obs{B}$ defined through $e_{\eqName{B}} = (\infty,0,\infty,0,0,\infty,\infty,\infty)$ equals the language defined by the grammar:
  \begin{align*}
    \varphi_\eqName{B} \quad ::= \quad & \hmlEps\hmlObs{a}\varphi_\eqName{B}\quad | \quad \hmlEps\hmlAndS \{\psi_\eqName{B}, \psi_\eqName{B}, \ldots\}
    \quad | \quad \hmlTrue \\
    \psi_\eqName{B} \quad ::= \quad    &
    \hmlNeg\hmlEps\hmlObs{a}\varphi_\eqName{B}
    \quad | \quad \hmlNeg\hmlEps\hmlAndS\{\psi_\eqName{B}, \psi_\eqName{B}, \ldots\}
    \quad | \quad \hmlEps\hmlObs{a}\varphi_\eqName{B}
    \quad | \quad \hmlEps\hmlAndS \{\psi_\eqName{B}, \psi_\eqName{B}, \ldots\}.
  \end{align*}
  Let us contrast this to the definition for weak bisimulation observations $\obs{B'}$ from Gazda et al.~\cite{gfm2020congruenceOperator}:
  \begin{align*}
    \varphi_\eqName{B'} \quad ::= \quad \hmlEps\varphi_\eqName{B'}
          \quad | \quad \hmlEps\hmlObs{a}\hmlEps\varphi_\eqName{B'}
          \quad | \quad \hmlAndS \{\varphi_\eqName{B'}, \varphi_\eqName{B'}, \ldots\} 
          \quad | \quad \hmlNeg\varphi_\eqName{B'}.
  \end{align*}
  Our $\obs{B}$ allows a few formulas that $\obs{B'}$ lacks, e.g.\@ $\hmlEps\hmlObs{a}\hmlEps\hmlObs{a}\hmlEps\hmlTrue$.
  This does not add expressiveness as $\obs{B'}$ has $\hmlEps\hmlObs{a}\hmlEps\hmlEps\hmlObs{a}\hmlEps\hmlTrue$ and $\hmlSemantics{\hmlEps\hmlEps\varphi}{}{} = \hmlSemantics{\hmlEps\varphi}{}{}$.

  For the other direction, there is a bigger difference due to $\obs{B'}$ allowing more freedom in the placement of conjunction and negation.
  In particular, it permits top-level conjunctions and negated conjunctions without $\hmlEps$ in between. But these features do not add distinctive power.
  $\obs{B'}$ also allows top-level negation, and this adds distinctive power to the preorders, effectively turning them into equivalence relations.
  We do not enforce this and thus our $\mathord{\bPreord{B}} \neq \mathord{\bEquiv{B}}$;
  e.g.\ $\tau \ccsPrefix \action a \bPreord{B} \tau \ccsChoice \tau \ccsPrefix \action a$, but $\tau \ccsChoice \tau \ccsPrefix \action a \not\bPreord{B} \tau \ccsPrefix \action a$ due to $\hmlEps\hmlAndS\{\hmlNeg\hmlEps\hmlObs{a}\hmlTrue\}$.
  However, as a distinction by $\hmlNeg \varphi$ in one direction implies one by $\varphi$ in the other, we know that this difference is ironed out once we consider the equivalence $\bEquiv{B}$.
\end{example}

\ifthenelse{\boolean{arxivversion}}{%
\begin{remark}
  None of the logics in \autoref{fig:ltbt-spectrum} restrict the first dimension,
  but the modal depth is kept to simplify the calculation of dimensions~6 and~7.
  It could also be used to define $k$-step bisimilarity and similar notions.

  More generally, there is no deeper necessity to use \emph{exactly} the dimensions that this paper employs or the original ones of~\cite{glabbeek1993ltbt}---in both cases, they are chosen in order to conveniently cover notions of equivalence that stem from varying contexts.
  To cover even more notions, additional dimensions would be necessary, as we will discuss in \autoref{sec:algo-refinements}.
\end{remark}
}{}

\section{A Game of Distinguishing Capabilities}
\label{sec:enrgy-game}

This section introduces a game to find out how two states can be distinguished in the silent-step spectrum:
Attacker tries to implicitly construct a distinguishing formula, defender wants to prove that no such formula exists.
The twist is that we use an \emph{energy} game where energies ensure the possible formulas to lie in sublogics along the lines of the previous section.

\subsection{Declining Energy Games}
\label{subsec:energy-games}

Equivalence problems of the strong linear-time--branching-time spectrum can be characterized as multi-dimensional declining energy games with special $\mathtt{min}$-operations between components as outlined in~\cite{bisping2023equivalenceEnergyGames}.
In this subsection, we revisit the definitions we will need in this paper.
For a more detailed presentation---in particular on how to compute attacker and defender winning budgets on this class of games---we refer to~\cite{bisping2023equivalenceEnergyGames} and~\cite{bg2023multiWeightedGames}.

\begin{definition}[Energy updates]
  \label{def:updates}
  The set of \emph{energy updates,} $\energyUpdates$, contains $\vectorComponents[8]{u} \in \energyUpdates$ where each component $u_k$ is a symbol of the form
  \begin{itemize}[noitemsep]
    \item $u_k \in \{-1, 0\}$ (relative update), or
    \item $u_k = \mathtt{min}_D$ where $D \subseteq \{1, \ldots, 8\}$ and $k \in D$ (minimum selection update).
  \end{itemize}

  \noindent
  Applying an update to an energy, $\energyUpdate(e, u)$, where $e = \vectorComponents[8]{e} \in \energies_\infty$ and $u = \vectorComponents[8]{u} \in \energyUpdates$, yields a new energy vector $e'$ where $k$th components $e'_k \defEq e_k + u_k$ for $u_k \in \ints$ and $e'_k \defEq \min_{d\in D} e_d$ for $u_k = \mathtt{min}_D$.
  Updates that would cause any component to become negative are undefined, i.e., $\energyUpdate$ is a partial function.
\end{definition}

\begin{example}
  $\energyUpdate((2,0,\infty,0,0,0,1,1), (\updMin{1,7},0,-1,0,0,0,0,-1))$ equals $(1,0,\infty,0,0,0,1,0)$.
\end{example}

\begin{definition}[Games]
  \label{def:energy-game}
  An $8$-dimensional \emph{declining energy game} $\game = (G, G_\defenderSubscript, \gameMove,\allowbreak w)$ is played on a directed graph uniquely labeled by energy updates consisting of
  \begin{itemize}[noitemsep]
    \item a set of \emph{game positions} $G$, partitioned into
    \begin{itemize}[noitemsep]
      \item \emph{defender positions} $G_\defenderSubscript\subseteq G$ and
      \item \emph{attacker positions} $G_\attackerSubscript\defEq G \setminus G_\defenderSubscript$,
    \end{itemize}
    \item a relation of \emph{game moves} $\operatorname{\gameMove}\subseteq G \times G$, and
    \item a \emph{weight function} for the moves $w \colon (\operatorname{\gameMove}) \to \energyUpdates$.
  \end{itemize}

  \noindent
  The notation $g \gameMoveX{u\,} g'$ stands for $g \gameMove g'$ and $w(g,g') = u$.
\end{definition}

\noindent
In the games of~\cite{bisping2023equivalenceEnergyGames}, the attacker wins precisely if they can get the defender stuck without running out of energy.
The energy budgets that suffice for the attacker to win from a game position can be characterized as follows:

\begin{definition}[Winning budgets]
  \label{def:win-characterization}
  The attacker winning budgets $\attackerWin^\game$ per position of a game $\game$ are defined inductively by the rules:
  \begin{mathparpagebreakable}
    \inferrule{
      g_\attackerSubscript \in G_\attackerSubscript\\
      g_\attackerSubscript \gameMoveX{u} g'\\
      \energyUpdate(e, u) \in \attackerWin^\game(g')
    } {
      e \in \attackerWin^\game(g_\attackerSubscript)
    }\and
    \inferrule{
      g_\defenderSubscript \in G_\defenderSubscript\\
      \forall u,g' \ldotp \; g_\defenderSubscript \gameMoveX{u} g' \longrightarrow
      \energyUpdate(e, u) \in \attackerWin^\game(g')
    } {
      e \in \attackerWin^\game(g_\defenderSubscript)
    }
  \end{mathparpagebreakable}
\end{definition}

\begin{figure*}[t]
  \centering
  \begin{adjustbox}{max width=\textwidth, center}
    \begin{tikzpicture}[>->,shorten <=1pt,shorten >=0.5pt,auto,node distance=2cm, rel/.style={dashed,font=\it},
      posStyle/.style={draw, inner sep=1ex,minimum size=1cm,minimum width=2cm,anchor=center,draw,black,fill=gray!5}]
        \node[posStyle, initial, initial text={}]
          (Att){$\attackerPos{p,Q}$};
        \node[posStyle]
          (AttDelay) [right = 1.6cm of Att] {$\attackerPos[\varepsilon]{p,Q_\varepsilon}$};
        \node[ellipse, draw, inner sep=1ex, minimum size=1cm,minimum width=2cm,fill=gray!5]
          (Def) [below = 2.2cm of AttDelay] {$\defenderPos{p,Q}$};
        \node[ellipse, draw=stabilityColor, inner sep=1ex, minimum size=1cm,minimum width=2cm,fill=stabilityColor!3]
          (DefStab) [above right = 1.2cm of Def] {$\defenderPosColor[s]{stabilityColor}{p,\{ q\!\in\!Q_\varepsilon \mid q \nostep{\tau} \}}$};
        \node[ellipse, draw=etaColor, inner sep=1ex, minimum size=1cm,minimum width=2cm,fill=etaColor!3]
          (DefBranch) [right = 3.3cm of AttDelay] {$\defenderPosColor[\eta]{etaColor}{p,\alpha,p',Q_\varepsilon \setminus Q_\alpha, Q_\alpha}$};
        \node[posStyle, draw=etaColor, fill=etaColor!3]
          (AttBranch) [above=.9cm of DefBranch] {$\attackerPosColor[\eta]{etaColor}{p',Q'}$};
        \node[posStyle]
          (AttConj) [below right = 1.2cm and .2cm of DefStab] {$\attackerPos[\wedge]{p,q}$};
        \node[posStyle, dashed, dash pattern={on 1.055045263157895mm off 0.993586315789474mm}, dash phase=0.405943684210525mm]
          (AttSwap) [right = 2cm of AttConj] {$\attackerPos[\varepsilon]{q,\{p' \mid p \stepWeak p'\}}$};
        \node[posStyle, dashed, dash pattern={on 1.051033684210526mm off 0.989808421052632mm}, dash phase=0.423411578947368mm]
          (AttContinue) [above = 1.5cm of AttSwap] {$\attackerPos[\varepsilon]{p,\{q' \mid q \stepWeak q'\}}$};
        \node[posStyle, dashed, dash pattern={on 1.03mm off 0.97mm}, dash phase=0.515mm]
          (AttObs) [right = 2cm of AttBranch] {$\attackerPos{p^\prime,Q^\prime}$};
        \draw[-] (AttSwap.south west) ++(0.2pt,0.2pt) ++(0.07,0) -- ++(-0.07,0) -- ++(0,0.05);
        \draw[-] (AttContinue.south west) ++(0.2pt,0.2pt) ++(0.07,0) -- ++(-0.07,0) -- ++(0,0.05);
        \draw[-] (AttObs.south west) ++(0.2pt,0.2pt) ++(0.07,0) -- ++(-0.07,0) -- ++(0,0.05);

        \path
          (Att) edge
            node[label={-90:$\textcolor{gray}{\zeroVec}$}] {$Q \stepWeak Q_\varepsilon$} (AttDelay)
          (Att) edge[bend right=15]
            node[pos = .4, label={-150:$\textcolor{gray}{\zeroVec}$}] {$Q = \varnothing$} (Def)
          (Att) edge[bend right=60]
            node[pos = .6, label={-130:$\textcolor{gray}{-\unit{5}}$}] {$Q \neq \varnothing$} (Def)
          (AttDelay) edge [out=155,in=115,looseness=4, pos=.6] node {$p \step{\tau} \ldots\; \textcolor{gray}{\zeroVec}$} (AttDelay)
          (AttDelay.north) edge[bend left=28]
            node[pos=.3, align=center, label={[label distance=0.0cm]-30:$\textcolor{gray}{-\unit{1}}$}] {$p\step{a}p'$\\ $Q_\varepsilon \step{a} Q'$} (AttObs)
          (AttDelay) edge
            node[pos =.3, label={[label distance=0cm]-180:$\textcolor{gray}{\zeroVec}$}] {$Q = Q_\varepsilon$} (Def)
          (AttDelay) edge[draw=stabilityColor]
            node[pos=.7,label={[label distance=0.2cm]-175:$\textcolor{stabilityColor!50}{\zeroVec}$}] {$\textcolor{stabilityColor}{p \centernot{\step{\tau}}}$} (DefStab)
          (AttDelay) edge[draw=etaColor]
            node[pos=.65,align=center,label={[label distance=0cm]-90:$\textcolor{etaColor!50}{\zeroVec}$}] {$\textcolor{etaColor}{p \step{\hmlOpt\alpha} p'}$\\$\textcolor{etaColor}{Q_\alpha \subseteq Q_\varepsilon}$} (DefBranch)
          (Def) edge[bend right=17]
            node[pos=.85, label={[label distance=0.1cm]-100:$\textcolor{gray}{-\unit{3}}$}] {$q \in Q$} (AttConj)
          (DefStab) edge[draw=stabilityColor, bend right=10]
            node[align=right, pos=.85, label={[label distance=0.3cm]-170:$\textcolor{stabilityColor!50}{-\unit{4}}$}] {\mbox{}\hspace*{-1.5em}$\textcolor{stabilityColor}{q \in Q_\varepsilon}$\\[2pt]\mbox{}\hspace*{-1.5em}$\textcolor{stabilityColor}{q \nostep{\tau}}$} (AttConj)
          (DefStab) edge[draw=stabilityColor, bend left=10]
            node[align=right, pos=.2, label={[label distance=0.3cm]165:$\textcolor{stabilityColor!50}{-\unit{4}}$}] {\mbox{}\hspace*{-3em}$\textcolor{stabilityColor}{\varnothing = Q = \mbox{}}$\\[2pt]\mbox{}\hspace*{-3em}$\textcolor{stabilityColor}{\{ q\!\in\!Q_\varepsilon \mid q \nostep{\tau} \}\;}$} (Def)
          (AttConj) edge[bend left=10]
            node[label={[label distance=0.0cm]-75:$\textcolor{gray}{\updMin{1,6},0,0,0,0,0,0,0}$}] {} (AttContinue)
          (AttConj) edge[bend right=15]
            node[below] {$\textcolor{gray}{\updMin{1,7},0,0,0,0,0,0,-1}$} (AttSwap)
          (DefBranch) edge[draw=etaColor]
            node[pos=.42, label={[label distance=0.1cm]-0:$\textcolor{etaColor!50}{\updMin{1,6},-1,-1,0,0,0,0,0}$}] {$\textcolor{etaColor}{Q_\alpha\! \step{(\alpha)} Q'}$} (AttBranch)
          (AttBranch) edge[draw=etaColor]
            node[below]{$\textcolor{etaColor!50}{-\unit{1}}$} (AttObs)
          (DefBranch) edge[bend left=5, draw=etaColor]
            node[pos=.07, align=center, label={[label distance=0.05cm]-180:$\textcolor{etaColor!50}{-\unit{2}-\unit{3}}$}] {$\textcolor{etaColor}{q \in Q_\varepsilon \setminus Q_\alpha}$} (AttConj);
    \end{tikzpicture}
  \end{adjustbox}
  \caption{
    Schematic spectroscopy game $\gameSpectroscopy$ of Definitions~\ref{def:spectroscopy-delay-game} (the black part),
    \textcolor{stabilityColor}{\ref{def:spectroscopy-stability-game}}~(with position $\defenderPosColor[s]{stabilityColor}{\cdots}$),
    and~\textcolor{etaColor}{\ref{def:spectroscopy-game-full}}~(with positions $\defenderPosColor[s]{stabilityColor}{\cdots}$, $\defenderPosColor[\eta]{etaColor}{\cdots}$ and $\attackerPosColor[\eta]{etaColor}{\cdots}$).}
  \label{fig:spec-game}
\end{figure*}

\subsection{Delaying Observations in the Spectroscopy Energy Game}
\label{subsec:delay-spectroscopy-game}

We begin with the part of the game that adds the concept of ``delayed'' attack positions to the ``strong'' spectroscopy game of~\cite{bisping2023equivalenceEnergyGames}.
It matches the black part of the $\hmlB$-grammar of \autoref{def:hml-branching}.
\autoref{fig:spec-game} gives a schematic overview of the game rules, where the game continues from the dashed nodes as from the initial node.
The colors differentiate the layers of following definitions and match the scheme of \autoref{def:hml-branching} and \autoref{fig:ltbt-spectrum}.

\begin{definition}[Spectroscopy delay game]
  \label{def:spectroscopy-delay-game}
  For a system $\system=(\proc,\actions,\linebreak[0] \mathord{\step{}})$,
  the \emph{spectroscopy delay energy game} $\game_\varepsilon^\system =(G,G_\defenderSubscript, \linebreak[0] \mathord{\gameMove}, w)$
  consists of

  \TabPositions{16em,19.3em}
  \begin{itemize}[noitemsep]
    \item \emph{attacker positions} \tab
      $\attackerPos{p,Q}$ \tab $\in G_\attackerSubscript$,
    \item \emph{attacker delayed positions} \tab
      $\attackerPos[\varepsilon]{p,Q}$ \tab $\in G_\attackerSubscript$,
    \item \emph{attacker conjunct positions} \tab
      $\attackerPos[\wedge]{p,q}$ \tab $\in G_\attackerSubscript$,
    \item \emph{defender conjunction positions} \tab
      $\defenderPos{p,Q}$ \tab $\in G_\defenderSubscript$,
  \end{itemize}

  \noindent
  where $p, q \in \proc$, $Q \in \powerSet{\proc}$, and nine kinds of moves:

  \TabPositions{8em, 11.4em, 19.65em,24.5em}
  \begin{itemize}[noitemsep]
    \item \emph{delay} \tab
      $\attackerPos{p,Q}$ \tab
      $\gameMoveX{\mathmakebox[6.3em]{0,0,0,0,0,0,0,0}}$ \tab
      $\attackerPos[\varepsilon]{p,Q'}$ \tab
      if $Q \stepWeak Q'$,
    \item \emph{procrastination} \tab
      $\attackerPos[\varepsilon]{p,Q}$ \tab
      $\gameMoveX{\mathmakebox[6.3em]{0,0,0,0,0,0,0,0}}$ \tab
      $\attackerPos[\varepsilon]{p',Q}$ \tab
      if $p \step{\tau} p'$, $p \neq p'$,
    \item \emph{observation} \tab
      $\attackerPos[\varepsilon]{p,Q}$ \tab
      $\gameMoveX{\mathmakebox[6.3em]{-1,0,0,0,0,0,0,0}}$ \tab
      $\attackerPos{p',Q'}$\tab
      if $p \step{a}p'$, $Q \step{a} Q' $, $a \neq \tau$,
    \item \emph{finishing} \tab
      $\attackerPos{p,\varnothing}$ \tab
      $\gameMoveX{\mathmakebox[6.3em]{0,0,0,0,0,0,0,0}}$ \tab
      $\defenderPos{p,\varnothing}$,
    \item \emph{immediate conj.} \tab
      $\attackerPos{p,Q}$ \tab
      $\gameMoveX{\mathmakebox[6.3em]{0,0,0,0,-1,0,0,0}}$ \tab
      $\defenderPos{p,Q}$ \tab
      if $Q \neq \varnothing$,
    \item \emph{late conj.} \tab
      $\attackerPos[\varepsilon]{p,Q}$ \tab
      $\gameMoveX{\mathmakebox[6.3em]{0,0,0,0,0,0,0,0}}$ \tab
      $\defenderPos{p,Q}$,
    \item \emph{conj.\ answer} \tab
      $\defenderPos{p,Q}$ \tab
      $\gameMoveX{\mathmakebox[6.3em]{0,0,-1,0,0,0,0,0}}$ \tab
      $\attackerPos[\wedge]{p,q}$ \tab
      if $q \in Q$,
    \item \emph{positive conjunct} \tab
      $\attackerPos[\wedge]{p,q}$ \tab
      $\gameMoveX{\mathmakebox[6.3em]{\updMin{1,6},0,0,0,0,0,0,0}}$ \tab
      $\attackerPos[\varepsilon]{p,Q}$ \tab
      if $\{q\} \stepWeak Q$,
    \item \emph{negative conjunct} \tab
      $\attackerPos[\wedge]{p,q}$ \tab
      $\gameMoveX{\mathmakebox[6.3em]{\updMin{1,7},0,0,0,0,0,0,-1}}$\tab
      $\attackerPos[\varepsilon]{q,Q}$\tab
      if $\{p\} \stepWeak Q$ and $p \neq q$.
  \end{itemize}
\end{definition}

\newcommand*{\seprule}{\\[-5pt]\rule{1.3cm}{.5pt}\\[-1pt]}
\begin{figure*}[t]
  \begin{adjustbox}{center}
    \begin{tikzpicture}[auto,shorten <=1pt,shorten >=0.5pt,
      defender/.style={ellipse, inner sep=0ex},
      defenderWins/.style={draw=blue, fill=blue!4, dashed, dash pattern=on 3.364pt off 3pt, dash phase=2.15pt},
      position/.style={inner sep=5pt,align=center,anchor=center,draw,black,fill=gray!5, node font=\small}]
      \begin{scope}[]
        \node[position, initial, initial text={}, label={100:$\color{magenta}\hmlEps\hmlAndS\{\hmlNeg\hmlEps\hmlObs{b}\}$}] (Ae_AB_A) at(0,0) {
          $\attackerPos[]{\ccsIdentifier{A^\tau_e}, \{ \ccsIdentifier{A^\tau_\ell}, \ccsIdentifier{B^\tau_\ell} \}}$
          \seprule
          \EqCoords{1,0,1,0,0,0,1,1}
        };

        \node[position, label={100:$\color{magenta}\hmlAndS\{\hmlNeg\hmlEps\hmlObs{b}\}$}] (Ae_AB_ADel) [right = 1.5cm of Ae_AB_A] {
          $\attackerPos[\varepsilon]{\ccsIdentifier{A^\tau_e}, \{ \ccsIdentifier{A^\tau_\ell}, \ccsIdentifier{B^\tau_\ell} \}}$
          \seprule
          \EqCoords{1,0,1,0,0,0,1,1}
        };
        \node[position, defenderWins] (End_End_ADel) [right = 1cm of Ae_AB_ADel] {
          $\attackerPosColor[]{blue}{\ccsStop, \{ \ccsStop \}}$};
        \node[position, defender, label={97:$\color{magenta}\hmlAndS\{\hmlNeg\hmlEps\hmlObs{b}\}$}] (Ae_AB_D) [below = 1.0cm of Ae_AB_ADel] {
          $\defenderPos[]{\ccsIdentifier{A^\tau_e}, \{ \ccsIdentifier{A^\tau_\ell}, \ccsIdentifier{B^\tau_\ell} \}}$
          \seprule
          \EqCoords{1,0,1,0,0,0,1,1}
        };
        \draw[blue] (End_End_ADel.south west) ++(0.2pt,0.2pt) ++(0.07,0) -- ++(-0.07,0) -- ++(0,0.05);

        \node[position, label={100:$\color{magenta}\hmlNeg\hmlEps\hmlObs{b}\hmlTrue$}] (Ae_B_AConj) [below left = .7cm and .1cm of Ae_AB_D] {
          $\attackerPos[\wedge]{\ccsIdentifier{A^\tau_e}, \ccsIdentifier{B^\tau_\ell}}$
          \seprule
          \EqCoords{1,0,0,0,0,0,1,1}
        };

        \node[position, label={80:$\color{magenta}\hmlNeg\hmlEps\hmlObs{b}\hmlTrue$}] (Ae_A_AConj) [below right = .7cm and .1cm of Ae_AB_D] {
          $\attackerPos[\wedge]{\ccsIdentifier{A^\tau_e}, \ccsIdentifier{A^\tau_\ell}}$
          \seprule
          \EqCoords{1,0,0,0,0,0,1,1}
        };

        \node[position, label={100:$\color{magenta}\hmlObs{b}\hmlTrue$}] (B_Ae_ADel) [below = 1cm of Ae_B_AConj] {
          $\attackerPos[\varepsilon]{\ccsIdentifier{B^\tau_\ell}, \{ \ccsIdentifier{A^\tau_e} \}}$
          \seprule
          \EqCoords{1,0,0,0,0,0,0,0}
        };

        \node[position, defender] (B_Ae_D) [left = .7cm of B_Ae_ADel] {
          $\defenderPos[]{\ccsIdentifier{B^\tau_\ell}, \{ \ccsIdentifier{A^\tau_e} \}}$
          \seprule
          \EqCoords{1,0,1,0,0,1,0,0}\\
          \EqCoords{1,0,2,0,0,0,1,2}
        };
        \node[position] (B_Ae_AConj) [above = .7cm of B_Ae_D] {
          $\attackerPos[\wedge]{\ccsIdentifier{B^\tau_\ell}, \ccsIdentifier{A^\tau_e} }$
          \seprule
          \EqCoords{1,0,0,0,0,1,0,0}\\
          \EqCoords{1,0,1,0,0,0,1,2}
        };

        \node[position, label={80:$\color{magenta}\hmlObs{b}\hmlTrue$}] (A_Ae_ADel) [below = 1cm of Ae_A_AConj] {
          $\attackerPos[\varepsilon]{\ccsIdentifier{A^\tau_\ell}, \{ \ccsIdentifier{A^\tau_e} \} }$
          \seprule
          \EqCoords{1,0,0,0,0,0,0,0}
        };

        \node[position, defender] (A_Ae_D) [right = .7cm of A_Ae_ADel] {
          $\defenderPos[]{\ccsIdentifier{A^\tau_\ell}, \{ \ccsIdentifier{A^\tau_e} \}}$
          \seprule
          \EqCoords{1,0,1,0,0,1,0,0}\\
          \EqCoords{1,0,2,0,0,0,1,2}
        };
        \node[position] (A_Ae_AConj) [above = .7cm of A_Ae_D] {
          $\attackerPos[\wedge]{\ccsIdentifier{A^\tau_\ell}, \ccsIdentifier{A^\tau_e} }$
          \seprule
          \EqCoords{1,0,0,0,0,1,0,0}\\
          \EqCoords{1,0,1,0,0,0,1,2}
        };

        \node[position, label={above left:$\color{magenta}\hmlTrue\vphantom{\hmlObs{b}}$}] (End_A) [below = 1cm of B_Ae_ADel] {
          $\attackerPos{\ccsStop, \varnothing}$
          \seprule
          \EqCoords{0,0,0,0,0,0,0,0}
        };
        \node[position, label={above left:$\color{magenta}\hmlTrue\vphantom{\hmlObs{b}}$}] (End_ADel) [right = 1cm of End_A] {
          $\attackerPos[\varepsilon]{\ccsStop, \varnothing}$
          \seprule
          \EqCoords{0,0,0,0,0,0,0,0}
        };
        \node[position, defender, label={above left:$\color{magenta}\hmlTrue$}] (End_D) [right = 1cm of End_ADel] {
          $\defenderPos{\ccsStop, \varnothing}$
          \seprule
          \EqCoords{0,0,0,0,0,0,0,0}
        };
      \end{scope}
      \begin{scope}[>->,black!75,every node/.style={node font=\small}]
        \draw[thick] (Ae_AB_A) to node {} (Ae_AB_ADel);
        \draw[thick] (Ae_AB_ADel) to node {} (Ae_AB_D);
        \draw (Ae_AB_ADel) to node {$-\unit{1}$} (End_End_ADel);
        \draw (Ae_AB_A.south) to[bend right=20, pos=.25] node {$-\unit{5}$} (Ae_AB_D.west);
        \draw[thick] (Ae_AB_D) to[bend left=15] node {$-\unit{3}$} (Ae_B_AConj);
        \draw[thick] (Ae_AB_D) to[bend right=15, swap] node {$-\unit{3}$} (Ae_A_AConj);
        \draw[thick] (Ae_B_AConj) to[pos=.5] node {$\updMin{1,7},...,-1$} (B_Ae_ADel);
        \draw[thick] (Ae_A_AConj) to[pos=.5, swap] node {$\updMin{1,7},...,-1$} (A_Ae_ADel);
        \draw[] (Ae_B_AConj) to[bend left=20, pos=.25] node {$\updMin{1,6}$} (Ae_AB_ADel);
        \draw[] (Ae_A_AConj) to[bend right=20, swap, pos=.25] node {$\updMin{1,6}$} (Ae_AB_ADel);
        \draw[] (B_Ae_ADel) to[bend left=5] node {} (A_Ae_ADel);
        \draw[thick] (A_Ae_ADel) to[bend left=5] node {} (B_Ae_ADel);
        \draw[] (B_Ae_ADel) to[bend left=5] node {} (B_Ae_D);
        \draw[thick] (B_Ae_D) to[bend left=5] node {$-\unit{3}$} (B_Ae_AConj);
        \draw[thick] (B_Ae_AConj) to[bend left=15, swap] node {$\updMin{1,6}$} (B_Ae_ADel.north west);
        \draw[thick] (B_Ae_AConj) to[bend left=5, pos=.3] node {$\updMin{1,7},...,-1$} (Ae_AB_ADel);
        \draw[] (A_Ae_ADel) to[bend right=5] node {} (A_Ae_D);
        \draw[thick] (A_Ae_D) to[bend right=5, swap] node {$-\unit{3}$} (A_Ae_AConj);
        \draw[thick] (A_Ae_AConj) to[bend right=15] node {$\updMin{1,6}$} (A_Ae_ADel.north east);
        \draw[thick] (A_Ae_AConj) to[bend right=5, pos=.3, swap] node {$\updMin{1,7},...,-1$} (Ae_AB_ADel);
        \draw[thick] (B_Ae_ADel) to node {$-\unit{1}$} (End_A);
        \draw[thick] (End_A) to node {} (End_ADel);
        \draw[thick] (End_A) to[bend right=20] node {} (End_D);
        \draw[thick] (End_ADel) to[swap] node {} (End_D);
      \end{scope}
    \end{tikzpicture}
  \end{adjustbox}
  \caption{Spectroscopy delay game $\game_\varepsilon$ from $\attackerPos{\ccsIdentifier{A^\tau_e}, \{ \ccsIdentifier{A^\tau_\ell}, \ccsIdentifier{B^\tau_\ell} \}}$ for \autoref{exa:attack-formula}. Each position names minimal attacker-winning budgets (due to the thick arrows) and corresponding distinguishing formulas (pink). Zeros and $\zeroVec$-updates are omitted for readability. Also, the game graph under defender-won reflexive position $\attackerPos{\ccsStop,\{ \ccsStop \}}$ (dashed in blue) is omitted.}
  \label{fig:example-game}
\end{figure*}

\begin{example}
  \label{exa:attack-formula}
  Starting at $\ccsIdentifier{P^\tau_e}$ and $\ccsIdentifier{P^\tau_\ell}$ of \autoref{exa:abstracted-processes} with energy $(2,0,1,0,0,0,1,1)$, the attacker can move with
  $\attackerPos{\ccsIdentifier{P^\tau_e}, \{ \ccsIdentifier{P^\tau_\ell} \}}
  \gameMoveX{\text{delay}}
  \gameMoveX{\text{observation\vphantom{y}}}
  \attackerPos{\ccsIdentifier{A^\tau_e}, \{ \ccsIdentifier{A^\tau_\ell}, \ccsIdentifier{B^\tau_\ell} \}}$.
  (For readability, we label the moves by the names of their rules.)
  This uses up $\unit{1}$ energy leading to level $(1,0,1,0,0,0,1,1)$.

  \autoref{fig:example-game} shows how the attacker can win from there.
  The attacker chooses a delay move and yields to the defender $\defenderPos[]{\ccsIdentifier{A^\tau_e}, \{ \ccsIdentifier{A^\tau_\ell}, \ccsIdentifier{B^\tau_\ell} \}}$.
  If the defender selects $\ccsIdentifier{B^\tau_\ell}$, bringing the energy to $(1,0,0,0,0,0,1,1)$, the attacker wins by
  $\attackerPos[\wedge]{
    \ccsIdentifier{A^\tau_e},
    \ccsIdentifier{B^\tau_\ell}
  }
  \gameMoveX{\text{negative conjunct}}
  \attackerPos[\varepsilon]{
    \ccsIdentifier{B^\tau_\ell},
    \ccsIdentifier{A^\tau_e}
  }
  \gameMoveX{\text{observation\vphantom{y}}}
  \gameMoveX{\text{finishing}}
  \mbox{$\defenderPos{\ccsStop, \varnothing}
  \centernot\gameMove$}$.
  For the defender choosing $\ccsIdentifier{A^\tau_\ell}$, a similar attack works due to
  $\attackerPos[\varepsilon]{
    \ccsIdentifier{A^\tau_\ell},
    \ccsIdentifier{A^\tau_e}
  }
  \gameMoveX{\text{procrastination}}
  \attackerPos[\varepsilon]{
    \ccsIdentifier{B^\tau_\ell},
    \ccsIdentifier{A^\tau_e}
  }$.
  Thus, the attacker wins the game.

  The tree of winning moves corresponds to formula $\varphi_\tau = \hmlEps\hmlObsI{op}\hmlEps\hmlAndS\{\hmlNeg\hmlEps\hmlObsI{b}\hmlTrue \}$ and budget of \autoref{exa:distinction-price}.
  This is no coincidence, but rather our core design principle for game moves.
  As we will prove in \autoref{sec:correctness}, attacker's winning moves match distinguishing $\hmlB$-formulas and their prices.

  Note that the attacker would not win if any component of the starting energy vector were lower.
  For example, $e_\mathrm{T} = (\infty,0,0,0,0,0,0,0) \notin \attackerWin(\attackerPos{\ccsIdentifier{P^\tau_e}, \{ \ccsIdentifier{P^\tau_\ell} \}})$ corresponds to weak trace inclusion, $\ccsIdentifier{P^\tau_e} \bPreord{T} \ccsIdentifier{P^\tau_\ell}$.
\end{example}

\subsection{Covering Stable Failures and Conjunctions}
\label{subsec:stable-game}

In order to cover ``stable'' and ``stability-respecting'' equivalences, we must separately count \textcolor{stabilityColor}{stable conjunctions}.

\begin{definition}[Spectroscopy stability game]
  \label{def:spectroscopy-stability-game}
  The \emph{stability game} $\game_s^\system$
  extends the delay game $\game_\varepsilon^\system$ of \autoref{def:spectroscopy-delay-game} by

  \begin{itemize}
    \item \emph{defender stable conjunction positions} \tabto{17em}
      $\defenderPos[s]{p,Q} \in G_\defenderSubscript$,
  \end{itemize}

  \noindent
  where $p \in \proc$, $Q \in \powerSet{\proc}$, and three kinds of moves:

  \TabPositions{9em, 12.4em,18.5em,22.5em}
  \begin{itemize}[noitemsep]
    \item \emph{stable conj.} \tab
      $\attackerPos[\varepsilon]{p,Q}$ \tab
      $\gameMoveX{\mathmakebox[5em]{0,0,0,0,0,0,0,0}}$ \tab
      $\defenderPos[s]{p,Q'}$ \tab
      if $Q' = \{ q \in Q \mid q \nostep{\tau} \}$,
      $p \nostep{\tau}$,
    \item \emph{conj. stable answer} \tab
      $\defenderPos[s]{p,Q}$ \tab
      $\gameMoveX{\mathmakebox[5em]{0,0,0,\textcolor{stabilityColor}{-1},0,0,0,0}}$ \tab
      $\attackerPos[\wedge]{p,q}$ \tab
      if $q \in Q$,
    \item \emph{stable finishing} \tab
      $\defenderPos[s]{p,\varnothing}$ \tab
      $\gameMoveX{\mathmakebox[5em]{0,0,0,\textcolor{stabilityColor}{-1},0,0,0,0}}$ \tab
      $\defenderPos{p,\varnothing}$.
  \end{itemize}
\end{definition}

\noindent
In principle, we add a move to enter a defender stable conjunction position and a move to leave it, similar to the defender conjunction positions in \autoref{def:spectroscopy-delay-game}.

\begin{example}
  Note that these new rules allow no new (incomparable) wins for the attacker in \autoref{exa:attack-formula}.
  Therefore, \emph{stable bisimulation} is another finest preorder (and equivalence) for the example processes
  because $e_\mathrm{SB} \notin \attackerWin(\attackerPos{\ccsIdentifier{P^\tau_e}, \{ \ccsIdentifier{P^\tau_\ell} \}})$ for $\game_s$.
\end{example}

\subsection{Extending to Branching Bisimulation}
\label{subsec:branching-game}

One last kind of distinctions is necessary to characterize \emph{branching bisimilarity,} the strongest common abstraction of bisimilarity for systems with silent steps:
its characteristic \textcolor{etaColor}{branching conjunctions}.

\begin{definition}[Weak spectroscopy game]
  \label{def:spectroscopy-game-full}
  The \emph{weak spectroscopy energy game} $\gameSpectroscopy^\system$
  extends \autoref{def:spectroscopy-stability-game} by

  \begin{itemize}[noitemsep]
    \item \emph{defender branching positions} \tabto{14em}
      $\defenderPos[\eta]{p,\alpha,p',Q,Q_\alpha} \in G_\defenderSubscript$,
    \item \emph{attacker branching positions} \tabto{14em}
      $\attackerPos[\eta]{p,Q} \in G_\attackerSubscript$,
  \end{itemize}

  \noindent
  where $p, p' \in \proc$ and $Q, Q_\alpha \in \powerSet{\proc}$ as well as $\alpha \in \actions$, and four kinds of moves:

  \TabPositions{9em,16.1em,24em,29em}
  \begin{itemize}[noitemsep]
    \item \emph{branching conj.} \tab
      $\attackerPos[\varepsilon]{p,Q}$\;
      $\gameMoveX{\mathmakebox[6em]{0,0,0,0,0,0,0,0}}$\;
      $\defenderPos[\eta]{p,\alpha,p',Q \setminus Q_\alpha,Q_\alpha}$\tab
      if $p \step{\hmlOpt\alpha} p'\!$, $Q_\alpha\subseteq Q$,
    \item \emph{branch.\ answer} \tab
      $\defenderPos[\eta]{p,\alpha,p'\!,Q,Q_\alpha}$\tab
      $\gameMoveX{\mathmakebox[7em]{0,\textcolor{etaColor}{-1},-1,0,0,0,0,0}}$\tab
      $\attackerPos[\wedge]{p,q}$\tab
      if $q \in Q$,
    \item \emph{branch.\ observation} \tab
      $\defenderPos[\eta]{p,\alpha,p'\!,Q,Q_\alpha}$\tab
      $\gameMoveX{\mathmakebox[7em]{\updMin{1,6},\textcolor{etaColor}{-1},-1,0,0,0,0,0}}$\tab
      $\attackerPos[\eta]{p',Q'}$\tab
      with $Q_\alpha\step{\hmlOpt\alpha}Q'\!$,
    \item \emph{branch.\ accounting} \tab
      $\attackerPos[\eta]{p,Q}$\;
      $\gameMoveX{\mathmakebox[10.8em]{-1,0,0,0,0,0,0,0}}$\tab
      $\attackerPos{p,Q}$.
  \end{itemize}
\end{definition}

\noindent
Intuitively, the attacker picks a step $p \step{\alpha} p'$ and some $Q_\alpha \subseteq Q$ that they claim to be inable to immediately simulate this step. For the remaining $Q \setminus Q_\alpha$, the attacker claims that these can be dealt with by other (possibly negative) delayed observations.
The defender then chooses which claim to counter.

\begin{example}
  Consider the CCS processes $\action{a} \ccsChoice \tau \ccsPrefix \action{b} \ccsChoice \action{b}$ and $\action{a} \ccsChoice \tau \ccsPrefix \action{b}$.
  The first process explicitly allows a $\action b$ to happen before deciding against $\action a$.
  To weak bisimilarity, for instance, this is transparent.
  To more branching-aware notions, it constitutes a difference.
  
  The two processes can be distinguished as follows in the weak spectroscopy game with energy budget $(1,\textcolor{etaColor}{1},1,0,0,1,0,0)$:
  First, the attacker enters a defender branching position $\attackerPos{\action{a} \ccsChoice \tau \ccsPrefix \action{b} \ccsChoice \action{b}, \{ \action{a} \ccsChoice \tau \ccsPrefix \action{b} \}}
  \gameMoveX{\text{delay}}
  \attackerPos[\varepsilon]{\action{a} \ccsChoice \tau \ccsPrefix \action{b} \ccsChoice \action{b}, \{ \action{a} \ccsChoice \tau \ccsPrefix \action{b}, \action{b} \}}
  \gameMoveX{\text{branching conjunction}}
  \defenderPos[\eta]{\action{a} \ccsChoice \tau \ccsPrefix \action{b} \ccsChoice \action{b}, \action{b}, \ccsStop,  \{ \action{b} \}, \{ \action{a} \ccsChoice \tau \ccsPrefix \action{b}\}}$.
  The defender can then pick between two losing options:
  \begin{itemize}[noitemsep]
      \item $\defenderPos[\eta]{\cdots}
        \gameMoveX{\text{branching answer}}
        \attackerPos[\wedge]{\action{a} \ccsChoice \tau \ccsPrefix \action{b} \ccsChoice \action{b}, \action{b}}$:
        Attacker responds
        $\attackerPos[\wedge]{\cdots}
        \gameMoveX{\text{positive conjunct}}
        \gameMoveX{\action{a}\text{-observation \vphantom{y}}}\linebreak[3]
        \gameMoveX{\text{finishing}}
        \defenderPos{\ccsStop, \varnothing}$,
        which corresponds to formula $\hmlEps\hmlObs{a}\hmlTrue$.
      \item $\defenderPos[\eta]{\cdots}
        \gameMoveX{\text{branching observation}}
        \attackerPos[\eta]{\ccsStop, \{\}}$:
        Attacker replies
        $\attackerPos[\eta]{\cdots}\allowbreak
        \gameMoveX{\text{branching accounting}}
        \gameMoveX{\text{finishing}}
        \defenderPos{\ccsStop, \varnothing}$,
        which corresponds to the $\hmlOpt{b}\hmlTrue$-ob\-ser\-va\-tion in the context of a branching conjunction. 
  \end{itemize}
  Taken together, the attacker wins this game constellation with a strategy that corresponds to the formula $\hmlEps\hmlAndS\{ \hmlOpt{b}, \hmlEps\hmlObs{a} \}$.

  The formula disproves $\eta$-simulation preorder and thus branching bisimilarity.
  However, the two processes are (stability-respecting) delay-bisimilar as there are no delay bisimulation formulas to distinguish them.
\end{example}

\section{Correctness}
\label{sec:correctness}

We now state in what sense winning energy levels and equivalences coincide in the context of a transition system $\system=(\proc,\actions,\linebreak[0] \mathord{\step{}})$.
\begin{theorem}[Correctness]
  For all $e \in \energies_\infty$, $p \in \proc$, $Q \in \powerSet{\proc}$, the following are equivalent:
  \begin{enumerate}[noitemsep]
    \item There exists a formula $\varphi \in \hmlB$ with price $\expr(\varphi) \leq e$ that distinguishes $p$ from $Q$.
    \item Attacker wins $\gameSpectroscopy^\system$ from $\attackerPos{p,Q}$ with $e$ (that is, $e \in \attackerWin^{\gameSpectroscopy^\system}(\attackerPos{p,Q})$).
  \end{enumerate}
\end{theorem}

\noindent
With \autoref{def:language-prices}, this means that,
for a notion of equivalence $N$ with coordinate $e_N$ in \autoref{fig:ltbt-spectrum},
$p \bPreord{\mathit{N}} q$ precisely if
the defender wins, $e_N \notin \attackerWin(\attackerPos{p,\{q\}})$.

The proof of the theorem is given through the following three lemmas.
The direction from (1) to (2) is covered by \autoref{thm:distinction-price-energies} when combined with the upward-closedness of attacker winning budgets.
From (2) to (1), the link is established through \emph{strategy formulas} by Lemmas~\ref{lem:win-implies-strat-formula} and~\ref{lem:strat-formula-distinguishes}.
The proofs
\ifthenelse{\not\boolean{arxivversion}}{
  can be found on arXiv~\cite{bj2023silentStepSpectroscopyArxiv} and
}{}
have also been formalized in an Isabelle/HOL theory.%
\footnote{The formalization can be found on \url{https://github.com/equivio/silent-step-spectroscopy}.}

\subsection{Distinguishing formulas imply attacker-winning budgets}

\begin{lemmaE}[][see full proof]
  \label{thm:distinction-price-energies}
  If $\varphi \in \hmlB$ distinguishes $p$ from $Q$, then $\expr(\varphi) \in \attackerWin(\attackerPos{p,Q})$.
\end{lemmaE}
\begin{proof}
  By mutual structural induction on $\varphi$, $\chi$, and $\psi$ with respect to the following claims:
  \begin{enumerate}[noitemsep]
    \item If $\varphi \in \hmlB$ distinguishes $p$ from $Q \neq \varnothing$, then $\expr(\varphi) \in \attackerWin(\attackerPos{p,Q})$;
    \item If $\chi$ distinguishes $p$ from $Q \neq \varnothing$ and $Q$ is closed under $\stepWeak$ (that is $Q \stepWeak Q$),
    then $\expr^\varepsilon(\chi) \in \attackerWin(\attackerPos[\varepsilon]{p,Q})$;
    \item If $\psi$ distinguishes $p$ from $q$,
    then $\expr^\wedge(\psi) \in \attackerWin(\attackerPos[\wedge]{p,q})$.
    \item If $\hmlAndS \Psi$ distinguishes $p$ from $Q \neq \varnothing$,
    then $\expr^\varepsilon(\hmlAndS \Psi) \in \attackerWin(\defenderPos{p,Q})$;
    \item If $\hmlAndS \{ \hmlNeg\hmlObs{\tau}\hmlTrue \} \cup \Psi$ distinguishes $p$ from $Q \neq \varnothing$
    and all the processes in $Q$ are stable,
    then \linebreak $\expr^\varepsilon(\hmlAndS \{ \hmlNeg\hmlObs{\tau}\hmlTrue \} \cup \Psi) \in \attackerWin(\defenderPos[s]{p,Q})$;
    \item If $\hmlAndS \{ \hmlOpt{\alpha}\varphi' \} \cup \Psi$ distinguishes $p$ from $Q$,
    then, for any $p \step{\hmlOpt\alpha} p' \in \hmlSemantics{\varphi'}{}{}$ and $Q_\alpha = Q \setminus \hmlSemantics{\hmlObs{\alpha}\varphi'}{}{}$, $\expr^\varepsilon(\hmlAndS \{ \hmlOpt{\alpha}\varphi'\! \} \linebreak[2] \cup \Psi) \in \attackerWin(\defenderPos[\eta]{p,\alpha,p'\!,\mbox{$Q \setminus Q_\alpha$}, Q_\alpha})$.
  \end{enumerate}
  \ifthenelse{\boolean{arxivversion}}{
  \begin{proofE}
  If $Q = \varnothing$, the lemma is very easy to prove.
  So let us assume that $Q \neq \varnothing$ for the rest.
  To get an inductive property, we actually prove the following property:
  \begin{enumerate}
    \item If $\varphi \in \hmlB$ distinguishes $p$ from $Q \neq \varnothing$, then $\expr(\varphi) \in \attackerWin(\attackerPos{p,Q})$;
    \item If $\chi$ distinguishes $p$ from $Q \neq \varnothing$ and $Q$ is closed under $\stepWeak$ (that is $Q \stepWeak Q$),
    then $\expr^\varepsilon(\chi) \in \attackerWin(\attackerPos[\varepsilon]{p,Q})$;
    \item If $\psi$ distinguishes $p$ from $q$,
    then $\expr^\wedge(\psi) \in \attackerWin(\attackerPos[\wedge]{p,q})$.
    \item If $\hmlAndS \Psi$ distinguishes $p$ from $Q \neq \varnothing$,
    then $\expr^\varepsilon(\hmlAndS \Psi) \in \attackerWin(\defenderPos{p,Q})$;
    \item If $\hmlAndS \{ \hmlNeg\hmlObs{\tau}\hmlTrue \} \cup \Psi$ distinguishes $p$ from $Q \neq \varnothing$
    and all the processes in $Q$ are stable,
    then \linebreak $\expr^\varepsilon(\hmlAndS \{ \hmlNeg\hmlObs{\tau}\hmlTrue \} \cup \Psi) \in \attackerWin(\defenderPos[s]{p,Q})$;
    \item If $\hmlAndS \{ \hmlOpt{\alpha}\varphi' \} \cup \Psi$ distinguishes $p$ from $Q$,
    then, for any $p \step{\hmlOpt\alpha} p' \in \hmlSemantics{\varphi'}{}{}$ and $Q_\alpha = Q \setminus \hmlSemantics{\hmlObs{\alpha}\varphi'}{}{}$, $\expr^\varepsilon(\hmlAndS \{ \hmlOpt{\alpha}\varphi'\! \} \linebreak[2] \cup \Psi) \in \attackerWin(\defenderPos[\eta]{p,\alpha,p'\!,\mbox{$Q \setminus Q_\alpha$}, Q_\alpha})$.
  \end{enumerate}

  \noindent
  We prove this by mutual induction over the structure of $\varphi$, $\chi$, and $\psi$.

  \begin{enumerate}
    \item Assume $\varphi$ distinguishes $p$ from $Q \neq \varnothing$.
      \begin{description}
        \item[$\varphi = \hmlEps\chi$:]
          That means that there exists $p \stepWeak p' \in \hmlSemantics{\chi}{}{}$
          and $Q' \cap \hmlSemantics{\chi}{}{} = \varnothing$ for $Q \stepWeak Q'$.
          Therefore, $\chi$ distinguishes $p'$ from $Q'$ and $Q' \stepWeak Q'$.
          By induction hypothesis we conclude that $\expr^\varepsilon(\chi) \in \attackerWin(\attackerPos[\varepsilon]{p',Q'})$.
      
          There are moves $\attackerPos{p,Q} \gameMoveX{\text{delay}}
          \attackerPos[\varepsilon]{p,Q'} \overset{\text{\smaller procrastination}}{\gameMove \cdots \gameMove}
          \attackerPos[\varepsilon]{p',Q'}$.
          Using \autoref{def:win-characterization} over these moves,
          we can conclude that $\expr^\varepsilon(\chi) \in \attackerWin(\attackerPos{p,Q})$.
          We get the result because $\expr(\varphi) = \expr^\varepsilon(\chi)$.
      
        \item[$\varphi = \hmlAndS \Psi$:]
          There is the move $\attackerPos{p,Q} \gameMoveX{\text{immediate conj.}} \defenderPos{p,Q}$.
          By induction hypothesis we conclude that $\expr^\varepsilon(\hmlAndS \Psi) \in \attackerWin(\defenderPos{p,Q})$.
          Using \autoref{def:win-characterization}
          we immediately get that $\expr(\varphi) = \expr^\varepsilon(\varphi) + \unit{5} \in \attackerWin(\attackerPos{p,Q})$.
      \end{description}

    \item Assume $\chi$ distinguishes $p$ from $Q$ (and $Q \stepWeak Q$).
      \begin{description}
        \item[$\chi = \hmlObs{a}\varphi'$:]
        That means that there exists $p' \in \hmlSemantics{\varphi'}{}{}$ such that $p \step{a} p'$.
        On the other hand, $Q' \cap \hmlSemantics{\varphi'}{}{} = \varnothing$, where $Q \step{a} Q'$,
        and therefore $\varphi'$ distinguishes $p'$ from $Q'$.

        Now there is the move $\attackerPos{p,Q} \gameMoveX{\text{observation}} \attackerPos{p',Q'}$,
        By induction hypothesis we conclude that $\expr(\varphi') \in \attackerWin(\attackerPos{p',Q'})$.
        Because we can calculate $\expr^\varepsilon(\hmlObs{a}\varphi') := \unit{1} + \expr(\varphi')$,
        we know $\energyUpdate(\expr^\varepsilon(\chi),-\unit{1}) = \expr(\varphi')$.
        With \autoref{def:win-characterization},
        we get $\expr^\varepsilon(\chi) \in \attackerWin(\attackerPos{p,Q})$.

      \item[$\chi = \hmlAndS \Psi$:]
        There is the move $\attackerPos[\varepsilon]{p,Q}\allowbreak \gameMoveX{\text{late conj.}} \defenderPos{p,Q}$;
        we use the proof for $\defenderPos{p,Q}$ that follows in (4) and \autoref{def:win-characterization}
        to then get $\expr^\varepsilon(\chi) \in \attackerWin(\attackerPos{p,Q})$.

      \item[$\chi = \hmlAndS \{ \hmlNeg\hmlObs{\tau}\hmlTrue \} \cup \Psi$:]
        There is the move $\attackerPos[\varepsilon]{p,Q}\allowbreak \gameMoveX{\text{late stable conj.}} \defenderPos[s]{p,Q'}$,
        where $Q' = \{ q \in Q \mid q \nostep{\tau} \}$.
        If $Q'$ is not empty, we argue as in the previous case using (5).

        If $Q'$ is empty, there is the move $\defenderPos[s]{p,Q'} = \defenderPos[s]{p,\varnothing} \allowbreak \gameMoveX{\text{stable finishing}} \defenderPos{p,\varnothing}$.
        The latter position is stuck, so $\attackerWin(\defenderPos{p,\varnothing}) = \energies_\infty$
        and by \autoref{def:win-characterization}, $e \in \attackerWin(\defenderPos[s]{p,\varnothing})$ for all $e \geq \unit{4}$.
        Because $\expr^\varepsilon(\chi) \geq \expr^\varepsilon(\hmlAndS \{ \hmlNeg\hmlObs{\tau}\hmlTrue \}) = \unit{4}$,
        we get the result.
      
      \item[$\chi = \hmlAndS \{ \hmlOpt{\alpha}\varphi' \} \cup \Psi$:]
        Note that there must exist $p \step{\hmlOpt{\alpha}} p' \in \hmlSemantics{\varphi'}{}{}$
        (otherwise $p \not\in \hmlSemantics{\hmlOpt{\alpha}\varphi'}{}{} \supseteq \hmlSemantics{\chi}{}{}$,
        so $\chi$ would not distinguish $p$ from anything).
        Pick such a $p'$,
        and set $Q_\alpha = Q \setminus \hmlSemantics{\hmlObs{\alpha}\varphi'}{}{}$.
        Then there is the move $\attackerPos[\varepsilon]{p,Q} \gameMoveX{\text{branching conj.}} \defenderPos[\eta]{p,\alpha,p', \mbox{$Q \setminus Q_\alpha$},Q_\alpha}$;
        so we can use the proof for $\defenderPos[\eta]{p,\alpha, p'\!, \mbox{$Q \setminus Q_\alpha$},Q_\alpha}$ that follows in (6) and \autoref{def:win-characterization}
        to get $\expr^\varepsilon(\chi) \in \attackerWin(\attackerPos[\varepsilon]{p,Q})$.
      \end{description}
    \item Assume $\psi$ distinguishes $p$ from $q$.
      \begin{description}
        \item[$\psi = \hmlEps\chi$:]
          That means that there exists $p \stepWeak p' \in \hmlSemantics{\chi}{}{}$
          and $Q' \cap \hmlSemantics{\chi}{}{} = \varnothing$ for $\{q\} \stepWeak Q'$.
          Therefore, $\chi$ distinguishes $p'$ from $Q'$ and $Q' \stepWeak Q'$.
          By induction hypothesis we conclude that $\expr^\varepsilon(\chi) \in \attackerWin(\attackerPos[\varepsilon]{p',Q'})$.
      
          Now there is a move sequence $\attackerPos[\wedge]{p,q} \gameMoveX{\text{positive conjunct}}
          \attackerPos[\varepsilon]{p,Q'} \overset{\text{\smaller procrastination}}{\gameMove \cdots \gameMove}
          \attackerPos[\varepsilon]{p',Q'}$.
          Using \autoref{def:win-characterization} over the procrastination moves,
          we can conclude that $\expr(\hmlEps\chi) = \expr^\varepsilon(\chi) \in \attackerWin(\attackerPos[\varepsilon]{p,Q'})$.
          Calculation shows
          $\energyUpdate(\expr^\wedge(\psi), (\updMin{1,6},0,0,0,0,0,0,0)) \geq \expr^\varepsilon(\chi)$,
          and this allows to apply \autoref{def:win-characterization} and get the result.

        \item[$\psi = \hmlNeg\hmlEps\chi$:]
          That means that there exists $q \stepWeak q' \in \hmlSemantics{\chi}{}{}$
          and $P' \cap \hmlSemantics{\chi}{}{} = \varnothing$ for $\{p\} \stepWeak P'$.
          Therefore, $\chi$ distinguishes $q'$ from $P'$ and $P' \stepWeak P'$.
          By induction hypothesis we conclude that $\expr^\varepsilon(\chi) \in \attackerWin(\attackerPos[\varepsilon]{q',P'})$.
          A similar calculation as in the previous case shows
          $\energyUpdate(\expr^\wedge(\psi), (\updMin{1,7},\linebreak[0]0,0,0,0,0,0,-1)) \geq \expr^\varepsilon(\chi)$,
          and this allows to apply \autoref{def:win-characterization} and get the result.
      \end{description}
      \item Assume $\hmlAndS \Psi$ distinguishes $p$ from $Q$.

        We can find, for every $q \in Q$, some $\psi_q \in \Psi$ such that $q \notin \hmlSemantics{\psi_q}{}{}$
        (so $\Psi \neq \varnothing$).
        Choose one such covering of $\psi_q$s.
        Let $\Psi' := \{ \psi_q \mid q \in Q \} \subseteq \Psi$.
        Each $\psi_q$ either has the form $\hmlEps\chi_q$ or $\hmlNeg\hmlEps\chi_q$.
        It must be the case that $p \in \hmlSemantics{\hmlAnd{q}{Q} \psi_q}{}{}$ and $Q \cap \hmlSemantics{\hmlAnd{q}{Q} \psi_q}{}{} = \varnothing$.

        Now there are the moves $\defenderPos{p,Q}
        \gameMoveX{\text{conj.\@ answer}} \attackerPos[\wedge]{p,q}$
        for all $q \in Q$.
        We have to show that
        $e_0 := \expr^\varepsilon(\chi) = \unit{3} + \sup \{ \expr^\wedge(\psi) \mid \psi \in \Psi \} \in \attackerWin(\defenderPos{p,Q})$.
        As $\attackerWin(\defenderPos{p,Q})$ is upwards-closed,
        we can restrict the supremum to $\Psi'$ instead of $\Psi$, so
        it suffices to prove that $\sup \{ \expr^\wedge(\psi) + \unit{3} \mid \psi \in \Psi' \} \in \attackerWin(\defenderPos{p,Q})$.
        Now, to show this using \autoref{def:win-characterization},
        we have to quantify over all game moves from $\defenderPos{p,Q}$, i.e.\@ over all conjunction answers,
        which lead to the positions $\attackerPos[\wedge]{p,q}$ for $q \in Q$.
        We have $\energyUpdate(e_0, -\unit{3}) \geq \expr^\wedge(\psi_q)$,
        and by induction hypothesis know $\expr^\wedge(\psi_q)\in \attackerWin(\attackerPos[\wedge]{p,q})$.
        Applying \autoref{def:win-characterization} immediately leads to the desired result.
      \item Assume $\hmlAndS \{ \hmlNeg\hmlObs{\tau}\hmlTrue \} \cup \Psi$ distinguishes $p$ from $Q \not= \varnothing$, where $p$ and the processes in $Q$ are stable.

        We choose $\psi_q$ (for every $q \in Q$) and $\Psi'$ as in the previous case.

        Now there are the moves $\defenderPos[s]{p,Q}
        \gameMoveX{\text{conj.\@ s-answer}} \attackerPos[\wedge]{p,q}$
        for all $q \in Q$.
        We have to show that
        $\expr^\varepsilon(\chi) = \unit{4} + \sup \{ \expr^\wedge(\psi) \mid \psi \in \Psi \} \in \attackerWin(\defenderPos[s]{p,Q})$.
        This proceeds exactly as in the previous case.

      \item Assume $\hmlAndS \{ \hmlOpt{\alpha}\varphi' \} \cup \Psi$ distinguishes $p$ from $Q$, $p \step{\hmlOpt\alpha} p' \in \hmlSemantics{\varphi'}{}{}$ and $Q_\alpha = Q \setminus \hmlSemantics{\hmlObs{\alpha}\varphi'}{}{}$.


        There are the moves $\defenderPos[\eta]{p,\alpha,p',\linebreak[0] \mbox{$Q \setminus Q_\alpha$}, Q_\alpha}
        \gameMoveX{\text{br.\ answer}} \attackerPos[\wedge]{p,q}$
        for all $q \in Q \setminus Q_\alpha$.
        Additionally, there are the moves $\defenderPos[\eta]{p,\alpha,p',\linebreak[0] \mbox{$Q \setminus Q_\alpha$},Q_\alpha}
        \gameMoveX{\text{br.\ observation}} \attackerPos[\eta]{p',Q'}
        \gameMoveX{\text{br.\ accounting}} \attackerPos{p',Q'}$ for $Q_\alpha \step{\hmlOpt{\alpha}} Q'$,
        and $\varphi'$ distinguishes $p'$ from $Q'$.

        We have to show that
        $e_0 := \expr^\varepsilon(\chi) = \unit{2} + \unit{3} + \sup \{ \linebreak[2] \expr^\wedge(\hmlEps\hmlOpt{\alpha}\varphi') \} \cup \{ \expr^\wedge(\psi) \mid \psi \in \Psi \} \in \attackerWin(\defenderPos[\eta]{p,\alpha,p',\linebreak[0] \mbox{$Q \setminus Q_\alpha$},Q_\alpha})$.
        For the branching answer moves, this proceeds exactly as in the previous cases.
        For the branching observation move,
        we have to show that $e_2 := \energyUpdate(\energyUpdate(e_0,\linebreak[1] (\updMin{1,6},-1,-1,0,0,0,0,0)),-\unit{1}) \in \attackerWin(\attackerPos{p',Q'})$.
        We know $e_2 \geq \expr(\varphi')$.
        Moreover, we know $\expr(\varphi') \in \attackerWin(\attackerPos{p'\!,Q'})$
        by induction hypothesis (or, trivially, if $Q' = \varnothing$). 
        This suffices to apply \autoref{def:win-characterization} and get the result.
        \qedhere
    \end{enumerate}
  \end{proofE}
  }{
    Full proof in report~\cite{bj2023silentStepSpectroscopyArxiv}.
  }
\end{proof}

\begin{figure}[ph]
  \begin{mathparpagebreakable}
    \inferrule*[left=delay]{
      \attackerPos{p,Q} \gameMoveX{u} \attackerPos[\varepsilon]{p,Q'}\\
      e' = \energyUpdate(e, u) \!\in \attackerWin(\attackerPos[\varepsilon]{p,Q'})\\
      \chi \in \hmlStrategies(\attackerPos[\varepsilon]{p,Q'}, e')
    } {
      \hmlEps\chi \in \hmlStrategies(\attackerPos{p,Q}, e)
    }\and
    \inferrule*[left=procr]{
      \attackerPos[\varepsilon]{p,Q} \gameMoveX{u} \attackerPos[\varepsilon]{p',Q}\\
      e' = \energyUpdate(e, u) \!\in \attackerWin(\attackerPos[\varepsilon]{p',Q})\\
      \chi \in \hmlStrategies(\attackerPos[\varepsilon]{p',Q}, e')
    } {
      \chi \in \hmlStrategies(\attackerPos[\varepsilon]{p,Q}, e)
    }\and
    \inferrule*[left=observation]{
      \attackerPos[\varepsilon]{p,Q} \gameMoveX{u} \attackerPos{p',Q'}\\
      e' = \energyUpdate(e, u) \!\in \attackerWin(\attackerPos{p',Q'})\\
      p \step{a} p'\\
      Q \step{a} Q'\\
      \varphi \in \hmlStrategies(\attackerPos{p',Q'}, e')
    } {
      \hmlObs{a}\varphi \in \hmlStrategies(\attackerPos[\varepsilon]{p,Q}, e)
    }\and
    \inferrule*[left=immediate conj]{
      \attackerPos{p,Q} \!\gameMoveX{u} \defenderPos{p,Q}\\
      e'\! = \!\energyUpdate(e, u)\! \in \attackerWin(\defenderPos{p,Q})\\
      \varphi \in \hmlStrategies(\defenderPos{p,Q}, e')\\
    } {
      \varphi \in \hmlStrategies(\attackerPos{p,Q}, e)
    }\and
    \inferrule*[left=late conj]{
      \attackerPos[\varepsilon]{p,Q} \!\gameMoveX{u} \defenderPos{p,Q}\\
      e'\! = \!\energyUpdate(e, u)\! \in \attackerWin(\defenderPos{p,Q})\\
      \chi \in \hmlStrategies(\defenderPos{p,Q}, e')
    } {
      \chi \in \hmlStrategies(\attackerPos[\varepsilon]{p,Q}, e)
    }\and
    \inferrule*[left=conj]{
      \defenderPos{p,Q} \!\gameMoveX{u_q}\! \attackerPos[\wedge]{p,q}\\
      \forall q \in Q \ldotp \;
      e_q \!=\! \energyUpdate(e, u_q) \!\in \attackerWin(\attackerPos[\wedge]{p,q}) \; \land \;
      \psi_q \in \hmlStrategies(\attackerPos[\wedge]{p,q}, e_q)
    } {
      \hmlAndS\{ \psi_q \mid q \in Q\} \in \hmlStrategies(\defenderPos{p,Q}, e)
    }\and
    \inferrule*[left=pos]{
      \attackerPos[\wedge]{p,q} \gameMoveX{u} \attackerPos[\varepsilon]{p,Q'}\\
      e' = \energyUpdate(e, u) \in \attackerWin(\attackerPos[\varepsilon]{p,Q'})\\
      \chi \in \hmlStrategies(\attackerPos[\varepsilon]{p,Q'}, e')
    } {
      \hmlEps\chi \in \hmlStrategies(\attackerPos[\wedge]{p,q}, e)
    }\and
    \inferrule*[left=neg]{
      \attackerPos[\wedge]{p,q} \gameMoveX{u} \attackerPos[\varepsilon]{q,P'}\\
      e' = \energyUpdate(e, u) \in \attackerWin(\attackerPos[\varepsilon]{q,P'})\\
      \chi \in \hmlStrategies(\attackerPos[\varepsilon]{q,P'}, e')
    } {
      \hmlNeg \hmlEps\chi \in \hmlStrategies(\attackerPos[\wedge]{p,q}, e)
    }\and
    \inferrule*[left={\textcolor{stabilityColor}{stable}}]{
      \attackerPos[\varepsilon]{p,Q} \!\gameMoveX{u} \defenderPos[s]{p,Q'}\\
      e'\! = \!\energyUpdate(e, u)\! \in \attackerWin(\defenderPos[s]{p,Q'})\\
      \chi \in \hmlStrategies(\defenderPos[s]{p,Q'}, e')
    } {
      \chi \in \hmlStrategies(\attackerPos[\varepsilon]{p,Q}, e)
    }\and
    \inferrule*[left={\textcolor{stabilityColor}{stable conj}}]{
      \defenderPos[s]{p,Q} \!\gameMoveX{u_q}\! \attackerPos[\wedge]{p,q}\\
      Q \neq \varnothing\\
      \forall q \in Q \ldotp \;
      e_q \!=\! \energyUpdate(e, u_q) \!\in \attackerWin(\attackerPos[\wedge]{p,q}) \; \land \;
      \psi_q \in \hmlStrategies(\attackerPos[\wedge]{p,q}, e_q)
    } {
      \hmlAndS \left(\{\hmlNeg\hmlObs{\tau}\hmlTrue \} \cup \{ \psi_q \mid q \in Q \}\right) \in \hmlStrategies(\defenderPos[s]{p,Q}, e)
    }\and
    \inferrule*[left={\textcolor{stabilityColor}{stable finish}}]{
      \defenderPos[s]{p,\varnothing} \!\gameMoveX{u}\! \defenderPos{p,\varnothing}\\
      e' \!=\! \energyUpdate(e, u) \!\in \attackerWin(\defenderPos{p,\varnothing})
    } {
      \hmlAndS \{\hmlNeg\hmlObs{\tau}\hmlTrue \} \in \hmlStrategies(\defenderPos[s]{p,Q}, e)
    }\and
    \inferrule*[left={\textcolor{etaColor}{branch}}]{
      \attackerPos[\varepsilon]{p,Q} \!\gameMoveX{u} \defenderPos[\eta]{p,\alpha,p',Q',Q_\alpha}\\
      e'\! = \!\energyUpdate(e, u) \in \attackerWin(\defenderPos[\eta]{p,\alpha,p',Q',Q_\alpha})\\
      \chi \in \hmlStrategies(\defenderPos[\eta]{p,\alpha,p',Q',Q_\alpha}, e')
    } {
      \chi \in \hmlStrategies(\attackerPos[\varepsilon]{p,Q}, e)
    }\and
    \inferrule*[left={\textcolor{etaColor}{branch conj}}]{
      g_\mathrm{d} = \defenderPos[\eta]{p,\alpha,p',Q,Q_\alpha} \!\gameMoveX{u_\alpha}\! \attackerPos[\eta]{p',Q'} \!\gameMoveX{u'_\alpha}\! \attackerPos{p',Q'}\\
      e_\alpha \!=\! \energyUpdate(\energyUpdate(e, u_\alpha), u'_\alpha) \in \attackerWin(\attackerPos{p',Q'})\\
      \varphi_\alpha \in \hmlStrategies(\attackerPos{p',Q'}, e_\alpha)\\
      \forall q \in Q \ldotp \;
      g_\mathrm{d} \!\gameMoveX{u_q}\! \attackerPos[\wedge]{p,q} \; \land \;
      e_q \!=\! \energyUpdate(e, u_q) \!\in \attackerWin(\attackerPos[\wedge]{p,q}) \; \land \;
      \psi_q \!\in \hmlStrategies(\attackerPos[\wedge]{p,q}, e_q)
    } {
      \hmlAndS\! \left(\{\hmlOpt{\alpha} \varphi_\alpha \} \cup \{ \psi_q \mid q \in Q \}\right) \in \hmlStrategies(\defenderPos[\eta]{p,\alpha,p'\!,Q,Q_\alpha}, e)
    }
  \end{mathparpagebreakable}
  \caption{Strategy formula constructions for \autoref{def:strategy-formulas}.}
  \label{fig:strategy-formulas}
\end{figure}

\subsection{Winning attacks imply cheap distinguishing formulas}

\begin{definition}[Strategy formulas]
  \label{def:strategy-formulas}
  The set of \emph{attacker strategy formulas} $\hmlStrategies$ for a $\gameSpectroscopy$-position with given energy level $e$ is derived from the sets of winning budgets, $\attackerWin$, inductively according to the rules in \autoref{fig:strategy-formulas}.
\end{definition}

\noindent
As an example how to read the above rules, \emph{procr} states that if there is a move $\attackerPos[\varepsilon]{p,Q} \gameMoveX{u} \attackerPos[\varepsilon]{p',Q}$
(based on \autoref{def:spectroscopy-delay-game}, this must be a procrastination move),
and the strategy formulas of the latter position contain $\chi$, then also the strategy formulas of the former position contain $\chi$.

\begin{lemmaE}[][see full proof]
  \label{lem:win-implies-strat-formula}
  If $e \in \attackerWin(\attackerPos{p,Q})$, then there is $\varphi \in \hmlStrategies(\attackerPos{p,Q}, e)$ with $\expr(\varphi) \leq e$.
\end{lemmaE}
\begin{proof}
  By induction over the structure of \autoref{def:win-characterization}.
  \ifthenelse{\boolean{arxivversion}}{
  \begin{proofE}
   We prove a more detailed result, namely:
  \begin{enumerate}
    \item If $e \in \attackerWin(\attackerPos{p,Q})$,
      then there is $\varphi \in \hmlStrategies(\attackerPos{p,Q}, e)$ with price \mbox{$\expr(\varphi) \leq e$};
    \item If $e \in \attackerWin(\attackerPos[\varepsilon]{p,Q})$,
      then there is $\chi \in \hmlStrategies(\attackerPos[\varepsilon]{p,Q}, e)$ with $\expr^\varepsilon(\chi) \leq e$;
    \item If $e \in \attackerWin(\attackerPos[\wedge]{p,q})$,
      then there is $\psi \in \hmlStrategies(\attackerPos[\wedge]{p,q}, e)$ with $\expr^\wedge(\psi) \leq e$.

    \item If $e \in \attackerWin(\defenderPos{p,Q})$,
      then there is $\hmlAndS \Psi \in \hmlStrategies(\defenderPos{p,Q})$ with $\expr^\varepsilon(\hmlAndS \Psi) \leq e$;
    \item If $e \in \attackerWin(\defenderPos[s]{p,Q})$,
      then there is $\hmlAndS \{ \hmlNeg\hmlObs{\tau}\hmlTrue \} \cup \Psi \in \hmlStrategies(\defenderPos[s]{p,Q})$ with price $\expr^\varepsilon(\hmlAndS \{ \hmlNeg\hmlObs{\tau}\hmlTrue \} \cup \Psi) \leq e$;
    \item If $e \in \attackerWin(\defenderPos[\eta]{p,\alpha,p',\mbox{$Q \setminus Q_\alpha$},Q_\alpha})$,
      then there is $\hmlAndS \{ \hmlObs{\alpha}\varphi' \} \cup \Psi \in \hmlStrategies(\defenderPos[\eta]{p,\alpha, p',\linebreak[0] \mbox{$Q \setminus Q_\alpha$},Q_\alpha})$ with price $\expr^\varepsilon(\hmlAndS \{ \hmlOpt{\alpha}\varphi' \} \cup \Psi) \leq e$.
  \end{enumerate}

  \noindent
  We induct over game positions $g$ and energies $e$ according to the inductive \autoref{def:win-characterization}.
  We distinguish cases depending on the kind of position.

  \begin{enumerate}
  \item
    Assume $e \in \attackerWin(\attackerPos{p,Q})$.
    This must be due to one of the following moves:
    \begin{description}
      \item[Delay move {$\attackerPos{p,Q} \gameMoveX{\zeroVec} \attackerPos[\varepsilon]{p,Q_\varepsilon}$}:]
    We know that $e = \energyUpdate(e,\allowbreak \zeroVec) \in \attackerWin(\attackerPos[\varepsilon]{p,Q_\varepsilon})$,
    so by induction hypothesis we know that there exists $\chi \in \hmlStrategies(\attackerPos[\varepsilon]{p,Q_\varepsilon},e)$
    and $\expr^\varepsilon(\chi) \leq e$.
    But then $\hmlEps\chi \in \hmlStrategies(\attackerPos{p,Q},e)$ by rule (delay) of \autoref{def:strategy-formulas}
    and $\expr(\hmlEps\chi) = \expr^\varepsilon(\chi) \leq e$.

      \item[Immediate conj.\ move $\attackerPos{p,Q} \gameMoveX{-\unit{5}} \defenderPos{p,Q}$:]
    It must hold that $e' = \energyUpdate(e,\allowbreak-\unit{5}) \in \attackerWin(\defenderPos{p,Q})$,
    so by induction hypothesis we know that there exists a conjunction $\hmlAndS \Psi \in \hmlStrategies(\defenderPos{p,Q},e')$
    and $\expr^\varepsilon(\hmlAndS \Psi) \leq e'$.
    But then,
    $\hmlAndS \Psi \in \hmlStrategies(\attackerPos{p,Q},e)$
    by rule (immediate conj) of \autoref{def:strategy-formulas},
    and $\expr(\hmlAndS \Psi) \leq e$.
    \end{description}

  \item
    Assume $e \in \attackerWin(\attackerPos[\varepsilon]{p,Q})$.
    This must be due to one of the following moves:
    \begin{description}
      \item[Procrastination move {$\attackerPos[\varepsilon]{p,Q} \gameMoveX{\zeroVec} \attackerPos[\varepsilon]{p',Q}$}:]
        We know $\energyUpdate(e,\zeroVec) = e \in \attackerWin(\attackerPos[\varepsilon]{p',Q})$.
        By induction hypothesis, there is $\chi \in \hmlStrategies(\attackerPos[\varepsilon]{p',Q},e)$
        and $\expr^\varepsilon(\chi) \leq e$;
        therefore, by rule (procr) of \autoref{def:strategy-formulas},
        $\chi \in \hmlStrategies(\attackerPos[\varepsilon]{p,Q},e)$.

      \item[Late (unstable) conjunction move {$\attackerPos[\varepsilon]{p,Q} \gameMoveX{\zeroVec} \defenderPos{p,Q}$}:]
        It must be the case that $e \in \attackerWin(\defenderPos{p,Q})$.
        By induction hypothesis there is $\hmlAndS \Psi \in \hmlStrategies(\defenderPos{p,Q}, e)$
        and $\expr^\varepsilon(\hmlAndS \Psi) \leq e$;
        therefore, by rule (late conj) of \autoref{def:strategy-formulas},
        $\hmlAndS \Psi \in \hmlStrategies(\attackerPos[\varepsilon]{p,Q},e)$.

      \item[Stable conjunction move {$\attackerPos[\varepsilon]{p,Q}\!\gameMoveX{\zeroVec}\!\defenderPos[s]{p,\{ q \!\in\! Q \!\mid\! \mbox{$q \!\centernot{\step{\tau}}$} \}}$}:]
        It must hold that $p$ is stable and $e \in \attackerWin(\linebreak[3]\defenderPos{p,\{q \in Q \mid \mbox{$q \centernot{\step{\tau}}$} \}})$.
        By induction hypothesis there is some formula $\hmlAndS \{ \hmlNeg\hmlObs{\tau}\hmlTrue \} \cup \Psi \in \hmlStrategies(\defenderPos[s]{p,\{ q \in Q \mid \mbox{$q \centernot{\step{\tau}}$} \}})$
        and $\expr^\varepsilon(\hmlAndS \{ \hmlNeg\hmlObs{\tau}\hmlTrue \} \cup \Psi) \leq e$;
        thus, by rule (stable) of \autoref{def:strategy-formulas},
        $\hmlAndS \{ \hmlNeg\hmlObs{\tau}\hmlTrue \} \cup \Psi \in \hmlStrategies(\attackerPos[\varepsilon]{p,Q},e)$.

      \item[Branch.\ conjunction move {$\attackerPos[\varepsilon]{p,Q} \gameMoveX{\zeroVec} \defenderPos{p,\alpha,p'\!,Q\setminus Q_\alpha, Q_\alpha}$}:]
        It must hold that $e \in \attackerWin(\defenderPos{p,\alpha,p'\!,\linebreak[1] Q\setminus Q_\alpha, Q_\alpha})$.
        By induction hypothesis there is a formula $\hmlAndS \{ \hmlOpt{\alpha}\varphi' \} \cup \Psi \in \hmlStrategies(\defenderPos{p,\alpha,p',\linebreak[1] Q\setminus Q_\alpha, Q_\alpha})$
        and $\expr^\varepsilon(\hmlAndS \{ \hmlOpt{\alpha}\varphi' \} \cup \Psi) \leq e$;
        therefore, by rule (branch) of \autoref{def:strategy-formulas},
        $\hmlAndS \{ \hmlOpt{\alpha}\varphi' \} \cup \Psi \in \hmlStrategies(\attackerPos[\varepsilon]{p,Q})$.
    \end{description}

  \item
    Assume $e \in \attackerWin(\attackerPos[\wedge]{p,q})$.
    This must be due to one of the following moves:
    \begin{description}
      \item[Positive conjunct {$\vphantom{\xrightarrow{\updMin{1,7}}} \attackerPos[\wedge]{p,q} \gameMoveX{\updMin{1,6},0,0,0,0,0,0,0} \attackerPos[\varepsilon]{p,\{ q' \mid q \stepWeak q' \}}$}:]
        It must hold that $e' := \energyUpdate(e,\linebreak[2](\updMin{1,6},\linebreak[2]0,0,0,0,0,0,0)) \in \attackerWin(\attackerPos[\varepsilon]{p,\{ q' \mid q \stepWeak q' \}})$.
        By induction hypothesis there is some formula $\chi \in \hmlStrategies(\attackerPos[\varepsilon]{p,\{ q' \mid q \stepWeak q' \}}, e')$
        and $\expr^\varepsilon(\chi) \leq e'$;
        therefore, by rule (pos) of \autoref{def:strategy-formulas},
        $\hmlEps\chi \in \hmlStrategies(\attackerPos[\wedge]{p,q}, e)$.

      \item[Negative conjunct {$\vphantom{\xrightarrow{\updMin{1,7}}} \attackerPos[\wedge]{p,q}\gameMoveX{\updMin{1,7},0,0,0,0,0,0,-1} \attackerPos[\varepsilon]{q, \{ p' \mid p \stepWeak p' \}}$}:]
        It holds that $e' := \energyUpdate(e,\linebreak[1](\updMin{1,7},\linebreak[2]0,0,0,0,0,0,-1)) \in \attackerWin(\attackerPos[\varepsilon]{q,\{ p' \mid p \stepWeak p' \}})$.
        By induction hypothesis there is some formula $\chi \in \hmlStrategies(\attackerPos[\varepsilon]{q,\{ p' \mid p \stepWeak p' \}}, e')$
        and $\expr^\varepsilon(\chi) \leq e'$;
        therefore, by rule (neg) of \autoref{def:strategy-formulas},
        $\hmlNeg\hmlEps\chi \in \hmlStrategies(\attackerPos[\wedge]{p,q}, e)$.
    \end{description}

  \item
    Assume $e \in \attackerWin(\defenderPos{p,Q})$.
    For each move $\defenderPos{p,Q} \gameMoveX{-\unit{3}} \attackerPos[\wedge]{p,q}$,
    it must hold that $e' := \energyUpdate(e,-\unit{3}) \in \attackerWin(\attackerPos[\wedge]{p,q})$,
    so by induction hypothesis there are $\psi_q \in \hmlStrategies(\attackerPos[\wedge]{p,q})$ with $\expr^\wedge(\psi_q) \leq e'$.
    Therefore, by rule (conj) of \autoref{def:strategy-formulas},
    $\hmlAnd{q}{Q} \psi_q \in \hmlStrategies(\defenderPos{p,Q}, e)$,
    and $\expr^\varepsilon(\hmlAnd{q}{Q} \psi_q) = \unit{3} + \sup \{ \expr^\wedge(\psi_q) \mid q \in Q \} \leq e$.

  \item
    Assume $e \in \attackerWin(\defenderPos[s]{p,Q})$.
    If $Q \not= \varnothing$, we can argue similar to the previous case.

    If $Q = \varnothing$, the only move is $\defenderPos[s]{p,Q} = \defenderPos[s]{p,\varnothing} \gameMoveX{-\unit{4}} \defenderPos{p,\varnothing}$.
    It must be the case that $\energyUpdate(e,-\unit{4}) \geq \zeroVec$, or equivalently, $e \geq \unit{4}$.
    But then we have $\expr^\varepsilon(\hmlAndS \{ \hmlNeg\hmlObs{\tau}\hmlTrue \}) = \unit{4} \leq e$ as required.

  \item
    Assume $e \in \attackerWin(\defenderPos[\eta]{p,\alpha,p', \mbox{$Q \setminus Q_\alpha$},Q_\alpha})$.
    Then there are moves $\defenderPos[\eta]{p,\alpha,p',\mbox{$Q \setminus Q_\alpha$}, \linebreak[1]Q_\alpha}\linebreak[1] \gameMoveX{-\unit{2}-\unit{3}} \attackerPos[\wedge]{p,q}$
    for every $q \in Q \setminus Q_\alpha$;
    it must be the case that $e' := \energyUpdate(e,-\unit{2}-\unit{3}) \in \attackerWin(\attackerPos[\wedge]{p,q})$,
    so by induction hypothesis there are formulas $\psi_q \in \hmlStrategies(\attackerPos[\wedge]{p,q})$ with $\expr^\wedge(\psi_q) \leq e'$.
    Also, for the moves $\defenderPos[\eta]{p,\!\alpha,\!p'\!,\! \mbox{$Q \setminus Q_\alpha$},\!Q_\alpha} \gameMoveX{\updMin{1,6},-1,-1,0,0,0,0,0} \attackerPos[\eta]{p',Q'} \gameMoveX{-\unit{1}} \attackerPos{p',Q'}$,
    it must be the case that $e'' := \energyUpdate(\energyUpdate(e,(\updMin{1,6},-1,-1,\allowbreak 0,0,0,0,0)),-\unit{1}) \in \attackerWin(\attackerPos{p',Q'})$,
    so by induction hypothesis there is some $\varphi' \in \hmlStrategies(\attackerPos{p',Q'},e'')$ with $\expr(\varphi') \leq e''$.

    Hence, by rule (branch conj) of \autoref{def:strategy-formulas},
    $\hmlAndS \{ \hmlOpt{\alpha}\varphi' \} \cup \{ \psi_q \mid q \in Q \setminus Q_\alpha \} \in \hmlStrategies(\defenderPos[\eta]{p,\alpha,p',\linebreak[0] \mbox{$Q \setminus Q_\alpha$},Q_\alpha}, e)$
    and $\expr^\varepsilon(\hmlAndS \{ \hmlOpt{\alpha}\varphi' \} \cup \{ \psi_q \mid q \in Q \setminus Q_\alpha \}) \leq e$.
    \qedhere
  \end{enumerate}
  \end{proofE}
  }{
    Full proof in report~\cite{bj2023silentStepSpectroscopyArxiv}.
  }
\end{proof}

\begin{lemmaE}[][see full proof]
  \label{lem:strat-formula-distinguishes}
  If $\varphi \in \hmlStrategies(\attackerPos{p,Q}, e)$,
  then $\varphi$ distinguishes $p$ from~$Q$.
\end{lemmaE}
\begin{proof}
  By induction over the derivation of $~\cdots \in \hmlStrategies(g, e)$
  according to \autoref{def:strategy-formulas}.
  \ifthenelse{\boolean{arxivversion}}{
  \begin{proofE}
    Again, to get an inductive property, we actually prove the following:
    \begin{enumerate}
      \item If $\varphi \in \hmlStrategies(\attackerPos{p,Q}, e)$,
        then $\varphi$ distinguishes $p$ from $Q$;
      \item If $\chi \in \hmlStrategies(\attackerPos[\varepsilon]{p,Q}, e)$ and $Q \stepWeak Q$,
        then $\hmlEps\chi$ distinguishes $p$ from $Q$;
      \item If $\psi \in \hmlStrategies(\attackerPos[\wedge]{p,q}, e)$,
        then $\psi$ distinguishes $p$ from $\{ q \}$.

      \item If $\hmlAndS \Psi \in \hmlStrategies(\defenderPos{p,Q}, e)$,
        then $\hmlAndS \Psi$ distinguishes $p$ from $Q$;
      \item If $\hmlAndS \{ \hmlNeg\hmlObs{\tau}\hmlTrue \} \cup \Psi \in \hmlStrategies(\defenderPos[s]{p,Q}, e)$
        and $p$ 
              is stable,
        then the stable conjunction $\hmlAndS \{ \hmlNeg\hmlObs{\tau}\hmlTrue \} \cup \Psi$ distinguishes $p$ from $Q$;
      \item If $\hmlAndS \{ \hmlOpt{\alpha}\varphi' \} \cup \Psi \in \hmlStrategies(\defenderPos[\eta]{p,\alpha,p',Q \setminus Q_\alpha,Q_\alpha}, e)$,
         $p \step{\hmlOpt\alpha} p'$ and $Q_\alpha \subseteq Q$,
        then the branching conjunction $\hmlAndS \{ \hmlOpt{\alpha}\varphi' \} \cup \Psi$ distinguishes $p$ from $Q$.
    \end{enumerate}

    \noindent
    We prove the result by induction over the derivation of $~\cdots \in \hmlStrategies(g, e)$
    according to \autoref{def:strategy-formulas}.

    \begin{enumerate}
      \item Assume $\varphi \in \hmlStrategies(\attackerPos{p,Q}, e)$.
        \begin{description}
          \item[Due to rule (delay) in \autoref{def:strategy-formulas}:]
            Then $\varphi = \hmlEps\chi$ and for $Q'$ with $Q \stepWeak Q'$
            we have $\chi \in \hmlStrategies(\attackerPos[\varepsilon]{p,Q'}, e)$.
            By induction hypothesis, $\hmlEps\chi$ distinguishes $p$ from $Q'$,
            but then it also distinguishes $p$ from $Q \subseteq Q'$.

          \item[Due to rule (immediate conj) in \autoref{def:strategy-formulas}:]
            (immediate conj) has premise $\attackerPos{p,Q} \gameMoveX{u} \defenderPos{p,Q}$,
            but this move can be a finishing move $\attackerPos{p,\varnothing} \gameMoveX{\zeroVec} \defenderPos{p,\varnothing}$
            or an immediate conjunction move $\attackerPos{p,Q} \gameMoveX{-\unit{5}} \defenderPos{p,Q}$ with $Q \neq \varnothing$.
            In either case, we have that $\varphi = \hmlAndS \Psi \in \hmlStrategies(\defenderPos{p,Q}, \energyUpdate(e, u))$.
            By induction hypothesis, $\hmlAndS \Psi$ distinguishes $p$ from $Q$,
            and this is exactly what we need to prove about $\varphi = \hmlAndS \Psi$.
        \end{description}

      \item Assume $\chi \in \hmlStrategies(\attackerPos[\varepsilon]{p,Q}, e)$ and $Q \stepWeak Q$.
        \begin{description}
          \item[Due to rule (procr) in \autoref{def:strategy-formulas}:]
          Then there is a step $p \stepWeak p'$ such that $\chi \in \hmlStrategies(\attackerPos[\varepsilon]{p',Q}, e)$.
          By induction hypothesis, we have that $\hmlEps\chi$ distinguishes $p'$ from $Q$,
          but then it also distinguishes $p$ from $Q$.
      
        \item[Due to rule (observation) in \autoref{def:strategy-formulas}:]
          Then $\chi = \hmlObs{a}\varphi$
          and there are $p \step{a} p'$ and $Q \step{a} Q'$ such that $\varphi \in \hmlStrategies(\attackerPos{p',Q'},\energyUpdate(e,-\unit{1}))$.
          By induction hypothesis we have that $\varphi$ distinguishes $p'$ from $Q'$.
          Therefore, $p \in \hmlSemantics{\chi}{}{} \subseteq \hmlSemantics{\hmlEps\chi}{}{}$.
          If there were some $q \in Q \cap \hmlSemantics{\hmlEps\chi}{}{}$,
          then we would have a path $q \stepWeak q' \step{a} q'' \in \hmlSemantics{\varphi}{}{}$.
          But $q' \in Q$ because $Q \stepWeak Q$ and therefore $q'' \in Q' \cap \hmlSemantics{\varphi}{}{} = \varnothing$.
          Contradiction!
          Therefore $\hmlEps\chi$ distinguishes $p$ from $Q$.
      
        \item[Due to rule (late conj) in \autoref{def:strategy-formulas}:]
          Then $\chi \in \hmlStrategies(\defenderPos{p,Q},e)$.
          By induction hypothesis, $\chi$ distinguishes $p$ from $Q$.
          As in the previous case,
          we use $Q \stepWeak Q$ to get that $\hmlEps\chi$ distinguishes $p$ from $Q$.
      
        \item[Due to rule (stable) in \autoref{def:strategy-formulas}:]
          Then $\chi = \hmlAndS \{ \hmlNeg\hmlObs{\tau}\hmlTrue \} \cup \Psi \in \hmlStrategies(\defenderPos[s]{p, \{ q \in Q \mid \mbox{$q \centernot{\step{\tau}}$} \}}, e)$.
          By induction hypothesis, $\chi$ distinguishes $p$ from the stable states in $Q$.
          Therefore, $p \in \hmlSemantics{\chi}{}{} \subseteq \hmlSemantics{\hmlEps\chi}{}{}$.
          unstable states do not satisfy $\hmlNeg\hmlOpt{\tau}\hmlTrue$,
          so if there were some unstable $q \in Q \cap \hmlSemantics{\hmlEps\chi}{}{}$,
          then we would have a path $q \stepWeak \mbox{$q' \centernot{\step{\tau}}$}$
          with $q' \in \hmlSemantics{\chi}{}{}$.
          But $q' \in Q$ because $Q \stepWeak Q$,
          so $q'$ cannot satisfy $\chi$ by induction hypothesis.
          Contradiction!
          Therefore $\hmlEps\chi$ distinguishes $p$ from all states in $Q$.
      
        \item[Due to rule (branch) in \autoref{def:strategy-formulas}:]
          Then $\chi \in \hmlStrategies(\defenderPos[\eta]{p,\alpha, p', \linebreak[0] \mbox{$Q \setminus Q_\alpha$},Q_\alpha})$
          (for some $p \step{\hmlOpt\alpha} p'$ and $Q_\alpha \subseteq Q$).
          By induction hypothesis, $\chi$ distinguishes $p$ from $Q$.
          As in the previous case,
          we use $Q \stepWeak Q$ to get that $\hmlEps\chi$ distinguishes $p$ from $Q$.
        \end{description}

      \item Assume $\psi \in \hmlStrategies(\attackerPos[\wedge]{p,q}, e)$.
        \begin{description}
          \item[Due to rule (pos) in \autoref{def:strategy-formulas}:]
            Then $\psi$ is of the form $\hmlEps\chi$ and $\chi \in \hmlStrategies(\attackerPos[\varepsilon]{p, Q'},\linebreak[1] \energyUpdate(e,\linebreak[1] (\updMin{1,6},0,0,0,0,0,0,0)))$ for $\{q\} \stepWeak Q'$.
            By induction hypothesis, $\hmlEps\chi$ distinguishes $p$ from $Q'$,
            and because $q \in Q'$, it also distinguishes $p$ from $q$.

          \item[Due to rule (neg) in \autoref{def:strategy-formulas}:]
            Then $\psi$ is of the form $\hmlNeg\hmlEps\chi$ and $\chi \in \hmlStrategies(\attackerPos[\varepsilon]{q,P'}, \linebreak[0] \energyUpdate(e,\linebreak[1] (\updMin{1,7},0,0,0,0,0,0,-1)))$ for $\{p\} \stepWeak P'$.
            By induction hypothesis, $\hmlEps\chi$ distinguishes $q$ from $P'$,
            and because $p \in P'$, its negation $\psi$ distinguishes $p$ from $q$.
        \end{description}

      \item Assume $\hmlAndS \Psi \in \hmlStrategies(\defenderPos{p,Q}, e)$.
        \begin{description}
          \item[Due to rule (conj) in \autoref{def:strategy-formulas}:]
            Then $\Psi$ can be written as $\{ \psi_q \mid q \in Q \}$,
            where each $\psi_q \in \hmlStrategies(\attackerPos[\wedge]{p,q},\linebreak[0] \energyUpdate(e, -\unit{3}))$.
            By induction hypothesis, $\psi_q$ distinguishes $p$ from $q$,
            so also $\hmlAndS \Psi$ distinguishes $p$ from $q$.
            Because this holds for every $q \in Q$,
            we have that $\hmlAndS \Psi$ distinguishes $p$ from $Q$.
        \end{description}

      \item Assume $\hmlAndS \{ \hmlNeg\hmlObs{\tau}\hmlTrue \} \cup \Psi \in \hmlStrategies(\defenderPos[s]{p,Q}, e)$ and $p$ is stable.
        \begin{description}
          \item[Due to rule (stable conj) in \autoref{def:strategy-formulas}:]
            Then $\Psi$ can be written $\{ \psi_q \mid q \in Q \}$,
            where $\psi_q \in \hmlStrategies(\attackerPos[\wedge]{p,q},\linebreak[0] \energyUpdate(e, -\unit{4}))$.
            By induction hypothesis, $\psi_q$ distinguishes $p$ from $q$.
            Because this holds for every $q \in Q$ and $p$ is stable,
            we have that $\hmlAndS \{ \hmlNeg\hmlObs{\tau}\hmlTrue \} \cup \Psi$ distinguishes $p$ from $Q$.
          \item[Due to rule (stable fin.) in \autoref{def:strategy-formulas}:]
            Then we must have $Q = \varnothing$ and $\Psi = \varnothing$.
            As $p$ is stable, it satisfies $\hmlAndS \{ \hmlNeg\hmlObs{\tau}\hmlTrue \}$,
            i.e.\@ the formula in $\hmlStrategies(\defenderPos[s]{p,Q}, e)$.
        \end{description}

      \item Assume $\hmlAndS \{ \hmlOpt{\alpha}\varphi' \} \cup \Psi \in \hmlStrategies(\defenderPos[\eta]{p,\alpha,p',Q \setminus Q_\alpha,Q_\alpha}, e)$,
         $p \step{\hmlOpt\alpha} p'$ and $Q_\alpha \subseteq Q$.
        \begin{description}
          \item[Due to rule (branch conj) in \autoref{def:strategy-formulas}:]
            Then $\Psi$ can be written as $\{ \psi_q \mid q \in Q \setminus Q_\alpha \}$,
            where $\psi_q \in \hmlStrategies(\attackerPos[\wedge]{p,q},\linebreak[0] \energyUpdate(e, -\unit{2}-\unit{3}))$.
            By induction hypothesis, $\psi_q$ distinguishes $p$ from $q$.
            Because this holds for every $q \in Q \setminus Q_\alpha$,
            we have that $\hmlAndS \Psi$ distinguishes $p$ from $Q \setminus Q_\alpha$.

            Moreover, there are moves $\defenderPos[\eta]{p,\!\alpha,\!p'\!,\!\mbox{$Q \setminus Q_\alpha$},\!Q_\alpha} \gameMove \attackerPos[\eta]{p',Q'} \gameMove \attackerPos{p',Q'}$
            where $p \step{\hmlOpt\alpha} p'$, $Q_\alpha \step{\hmlOpt{\alpha}} Q'$,
            and $\varphi' \in \hmlStrategies(\attackerPos{p,Q}, \energyUpdate(\energyUpdate(e,\allowbreak(\updMin{1,6},0,0,0,0,0,0,0)),-\unit{1}))$.
            By induction hypothesis, $\varphi'$ dis\-tin\-gu\-ish\-es $p'$ from $Q'$,
            so $\hmlOpt{\alpha}\varphi'$ dis\-tin\-gu\-ish\-es $p$ from $Q_\alpha$.

            Together we have that $\hmlAndS \{ \hmlOpt{\alpha}\varphi' \} \cup \Psi$ distinguishes $p$ from $Q$.
            \qedhere
        \end{description}
    \end{enumerate}
  \end{proofE}
  }{
    Full proof in report~\cite{bj2023silentStepSpectroscopyArxiv}.
  }
\end{proof}

\begin{figure}
  \includegraphics[width=\textwidth]{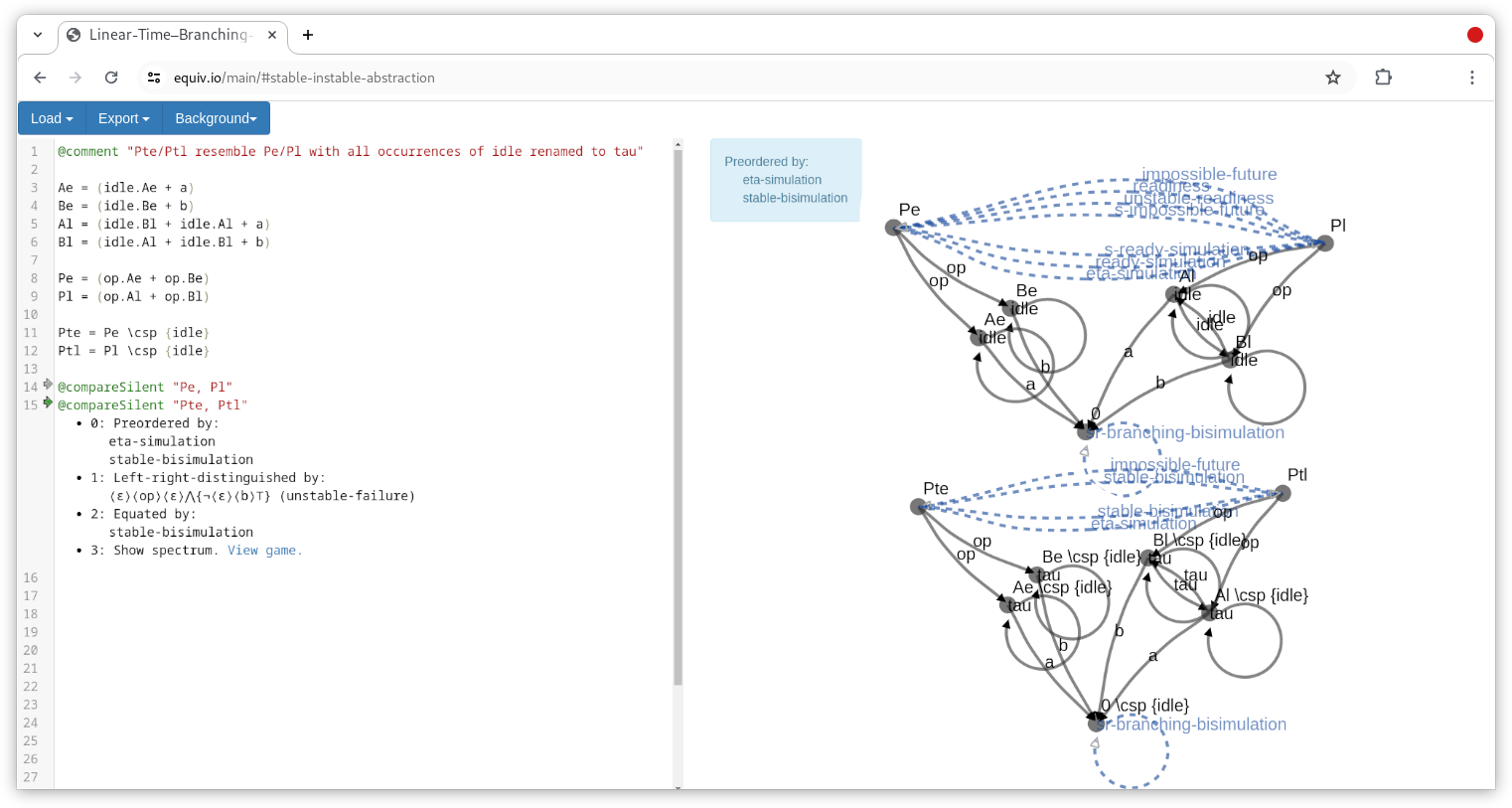}
  \caption{Screenshot of \url{equiv.io} solving \autoref{exa:distinctions-tool}.}
  \label{fig:equivio-screen}
\end{figure}

\section{Deciding All Weak Equivalences at Once}
\label{sec:algo-refinements}

The weak spectroscopy energy game enables algorithms to decide all considered behavioral equivalences.
An open-source prototype implementation 
can be tried out on \url{https://equiv.io}. 
Moreover, there is an extension of CAAL (Concurrency Workbench, Aalborg Edition,~\cite{aaehlosw2015caal}) with the entailed algorithm on~\url{https://github.com/equivio/CAAL}.
Both yield the expected output on the finitary examples from~\cite{glabbeek1993ltbt}.

The game allows \emph{checking individual equivalences} by instantiating it to start with an energy vector $e_N$ from \autoref{fig:ltbt-spectrum}.
The remaining reachability game can be decided with (usually exponential) time and space complexities depending on the selected energy vector.

\nopagebreak
More generally, one can \emph{decide all equivalences at once} by computing the pareto frontier of attacker budgets $\attackerWin(\attackerPos{p, \{q\}})$.
The algorithm of~\cite{bg2023multiWeightedGames} for multi-weighted games, has space complexity $\bigo(\relSize{G})$ and time complexity $\bigo(\relSize{\gameMove} \cdot \relSize{G} \cdot o)$ for bounded energies (due to a concrete spectrum),
where $o$ is the out-degree of $\gameMove$.
For this paper's weak spectroscopy game, $\gameSpectroscopy$,
we have
$\relSize{G_\vartriangle} \in \bigo (\relSize{\,\step{}} \cdot 3^{\relSize{\proc}})$
and $\relSize{\gameMove_\vartriangle} \in \bigo (\relSize{\,\step{}} \cdot \relSize{\proc} \cdot 3^{\relSize{\proc}})$,
and also $o_\vartriangle \in \bigo (\relSize{\,\step{}} \cdot 2^{\relSize{\proc}})$,
because of the defender branching positions and their surroundings.
This amounts to exponential time complexity.
Clearly, the approach is mostly tailored towards small examples.
But often these are all one needs:

\begin{example}
  \label{exa:distinctions-tool}
  Let us try our initial \autoref{exa:abstracted-processes} of abstracted processes
  (\autoref{fig:equivio-screen} and \url{https://equiv.io/#stable-unstable-abstraction}).
  The browser tool takes about 100~ms (considering a game of 112~positions) to report that $\ccsIdentifier{P_e}$ and $\ccsIdentifier{P_\ell}$ are stable \emph{and} unstable readiness-equivalent.
  $\ccsIdentifier{P^\tau_e}$ and $\ccsIdentifier{P^\tau_\ell}$ on the other hand are stable-bisimilar.
  This output immediately tells us that only notions either strictly finer than readiness or coarser than stable bisimilarity can be congruences for abstraction.
  In particular, unstable failures, which Gazda et al.~\cite[Corr.~9]{gfm2020congruenceOperator} report to be a congruence for abstraction, cannot be one because the unstable failure formula $\hmlEps\hmlObsI{op}\hmlEps\hmlAndS\{\hmlNeg\hmlEps\hmlObsI{a}\hmlTrue \}$ distinguishes $\ccsIdentifier{P^\tau_e}$ from $\ccsIdentifier{P^\tau_\ell}$, analogously to $\varphi_\tau$ of \autoref{exa:distinguishing-formula}.
\end{example}

\section{Related Work and Conclusion}
\label{sec:conclusion}

This paper provides the first \emph{generalized game characterization} for the spectrum of \emph{``weak'' behavioral equivalences} and preorders.
To this end, \autoref{sec:background} introduced a new \emph{modal characterization of branching bisimilarity}
that can be used to capture the \emph{modal logics of the silent-step spectrum}.
With this perspective, the set of weak equivalence problems becomes just one \emph{quantitative problem}, expressible as one energy game in \autoref{sec:enrgy-game}.

Other \emph{generalized game characterizations} by Chen and Deng~\cite{cd2008gameCharacetrizations} and by us~\cite{bjn2022decidingAllBehavioralEqs,bisping2023equivalenceEnergyGames} have only addressed strong equivalences
or parts of the spectrum~\cite{shr1995hornsatGames,tan2002abstractEquivalenceGames}.
Fahrenberg et al.~\cite{fahrenberg2014quantitativeLTBTS} treated a quantitative game interpretation for behavioral distances, as well disregarding silent-step notions.
Extending this line of work to account for silent steps in full is necessary for virtually every application.

In the silent-step spectrum, many things are more complicated.
There are \emph{several abstractions of bisimilarity}:
branching, $\eta$, delay and weak bisimilarity, as well as
contrasimilarity, stable bisimilarity and coupled similarity.
We have had to radically depart from their existing games \cite{ekw2017gamesBisimAbstraction,bnp2020coupledsim30,bm2021contrasimilarity} to cover all equivalences.
Depending on \emph{whether stabilization is required} for negated and conjunct observations, each equivalence notion has different weak versions.
Our game characterization is the first to explicitly consider stability-respecting notions, thereby unifying stable equivalences~\cite{glabbeek1993ltbt} and unstable ones~\cite{gfm2020congruenceOperator}.
This unification enables observations about the applicability of (un)stable equivalences as the one in \autoref{exa:distinctions-tool}.

The \emph{framework of codesigning games and grammars} can also easily be extended to cater for more notions, for instance, divergence-aware ones, or even to combine strong and weak ones in one game.
The connection to energy games enabled us to boost our approach using Brihaye and Goeminne's recent polynomial decision procedure for multi-weighted games~\cite{bg2023multiWeightedGames}.

We have added to the rich body of work on \emph{modal characterizations of branching bisimilarity}~\cite{nicolaVaandrager1995threeLogicsBB,glabbeek1993ltbt,FokkinkGL19DivCong3,geuversGolov2023positiveBB,geuvers2022}.
Continuing~\cite{bjn2022decidingAllBehavioralEqs,bisping2023equivalenceEnergyGames},
our work participates in a recent trend towards a modal focus for equivalences, also found
in Ford et al.~\cite{fms2021behavioralPreordGradedMonads} connecting graded modal logics and monads, and
in Wißmann et al.~\cite{wms2021explainingBehavioralInequivGenerically} as well as Beohar et al.~\cite{bgkm2023hmThmsViaGalois}.
Like Martens and Groote~\cite{mg2024minDepthDistBranchBisim}, we find minimal-depth distinguishing formulas for branching bisimilarity, but we solve the problem for all weak notions at once.

Our main related work, of course, is van Glabbeek's \emph{linear-time--branching-time spectrum}~\cite{glabbeek1990ltbt1,glabbeek1993ltbt}.
Up to today, part~II on silent steps is available only as “extended abstract” (in two versions!),
while part~I has seen a journal version~\cite{glabbeek2001ltbtsiReport} and refinements by others~\cite{erph2013unfyingLTBTS}.
We hope the present work makes the wisdom on weak equivalences of part~II more accessible to tools and humans alike.

{\small\paragraph*{Acknowledgments.}
We would like to thank Rob van Glabbeek and the EXPRESS/SOS'24 audience for discussing the material with us,
as well as several anonymous referees for pointing out weaknesses in a previous version of this paper.
Special thanks is due to the TU~Berlin students Lisa A. Barthel, Leonard M. Hübner, Caroline Lemke, Karl P. P. Mattes, and Lenard Mollenkopf, who validated the present paper in Isabelle/HOL, uncovering and addressing several flaws.
}

\bibliographystyle{eptcs}
\bibliography{bib}

\ifthenelse{\boolean{arxivversion}}{%

\appendix

\section{Proofs}

An Isabelle/HOL formalization of game and proofs can be found on \url{https://github.com/equivio/silent-step-spectroscopy}.

\printProofs
}{}

\end{document}

%% file: macros.tex
\global\long\def\defEq{\mathrel{\coloneqq}}%
\global\long\def\codeStyle#1{\mathrm{#1}}%
\global\long\def\varname#1{\mathsf{#1}}%

\newcommand*{\ccsRes}[1]{\left(\boldsymbol{\nu}#1\right)}%
\newcommand*{\ccsStop}{\mathsf{0}}%
\newcommand*{\ccsPrefix}{\ldotp\!}%
\newcommand*{\ccsChoice}{+}%
\newcommand*{\ccsDef}{\mathrel{\stackrel{\mathrm{def}}{=}}}%
\newcommand*{\ccsHide}{\mathrel{\backslash}}%
\newcommand*{\ccsPar}{\mathrel{\mid}}%
\newcommand*{\ccsRepl}{\textbf{!}}%
\newcommand*{\ccs}{\mathsf{CCS}}%
\newcommand*{\ccsValuation}{\mathcal{D}}%
\newcommand*{\ccsIdentifier}[1]{\mathsf{#1}}%
\newcommand*{\ccsOutm}[1]{\overline{#1}}%
\newcommand*{\ccsInm}[1]{#1}%

\global\long\def\rel#1{\mathcal{#1}}%
\global\long\def\bigo{\mathcal{O}}%
\newcommand*{\relSize}[1]{\lvert\mathord{#1}\rvert}
\newcommand{\powerSet}[1]{\mathbf{2}^{#1}}
\newcommand{\bellNumber}[1]{\mathbf{B}(#1)}
\newcommand{\compose}{\mathbin{\circ}}
\newcommand{\nats}{\mathbb{N}}
\newcommand{\ints}{\mathbb{Z}}
\newcommand{\domain}{\operatorname{\mathrm{dom}}}
\newcommand*{\vectorComponents}[2][n]{({#2}_1,\ldots,{#2}_{#1})}
\newcommand*{\Min}{\operatorname{\mathrm{Min}}}
\newcommand*{\Max}{\operatorname{\mathrm{Max}}}
\newcommand*{\unit}[1]{\hat{\mathbf{e}}_{#1}}
\newcommand*{\zeroVec}{\mathbf{0}}
\newcommand*{\formulaHighlight}[1]{\textcolor{violet}{#1}}

\newcommand*{\gameMoveX}[1]{\mathrel{\smash{\xrightarrowtail{\scriptscriptstyle#1}}}}%
\newcommand*{\gameMove}{\gameMoveX{\hspace*{0.5em}}}%
\newcommand*{\game}{\mathcal{G}}%
\newcommand*{\gameSpectroscopy}{\mathcal{G}_\vartriangle}%
\newcommand*{\gameSpectroscopyClever}{\mathcal{G}_\blacktriangle}%
\newcommand*{\attackerPos}[2][]{\textcolor{gray}{[}#2\textcolor{gray}{]_\mathtt{a}^{\smash{\scriptscriptstyle#1}}}}
\newcommand*{\defenderPos}[2][]{\textcolor{gray}{(}#2\textcolor{gray}{)_\mathtt{d}^{\smash{\scriptscriptstyle#1}}}}
\newcommand*{\attackerPosColor}[3][]{\textcolor{#2!50}{[}\textcolor{#2}{#3}\textcolor{#2!50}{]_\mathtt{a}^{\smash{\scriptscriptstyle#1}}}}
\newcommand*{\defenderPosColor}[3][]{\textcolor{#2!50}{(}\textcolor{#2}{#3}\textcolor{#2!50}{)_\mathtt{d}^{\smash{\scriptscriptstyle#1}}}}
\newcommand*{\partition}[1]{\mathscr{#1}}
\newcommand*{\conjClosure}[1]{\lceil #1 \rceil^\land}
\newcommand*{\attackerSubscript}{{\operatorname{a}}}
\newcommand*{\defenderSubscript}{{\operatorname{d}}}
\newcommand*{\energies}{\mathbf{En}}
\newcommand*{\energyUpdates}{\mathbf{Up}}%
\newcommand*{\energyUpdate}{\mathsf{upd}}%
\newcommand*{\energyUpdateInv}{\mathsf{upd}^{-1}}%
\newcommand*{\updMin}[1]{\mathtt{min}_{\{\!#1\!\}}}
\newcommand*{\energyLevel}{\mathsf{EL}}%
\newcommand*{\attackerWin}{\mathsf{Win}_\attackerSubscript}
\newcommand*{\attackerWinMin}{\attackerWin^{\scriptscriptstyle\min}}
\newcommand*{\defenderWinMax}{\mathsf{Win}_\defenderSubscript^{\scriptscriptstyle\max}}

\newcommand*{\proc}{\mathcal{P}}
\newcommand*{\system}{\mathcal{S}}%
\newcommand*{\action}[1]{\mathit{#1}}
\newcommand*{\actions}{\Sigma}%
\newcommand*{\step}[1]{\mathrel{\xrightarrow{\smash{#1}}}}%
\newcommand*{\nostep}[1]{\mathrel{\smash{\centernot{\xrightarrow{#1}}}}}%
\newcommand*{\stepWeak}{\mathrel{\twoheadrightarrow}}%
\newcommand*{\stepWord}[1]{\mathrel{\overset{\smash{\raisebox{-0.35ex}{$\scriptstyle #1$}}}{\twoheadrightarrow}}}%
\newcommand*{\initials}{\mathcal{I}}

\newcommand{\hml}{\mathsf{HML}}
\newcommand{\hmlA}{\hml[\actions]}
\newcommand{\hmlB}{\hml_{\mathrm{srbb}}}
\newcommand{\hmlBA}{\hmlB[\actions]}
\newcommand{\hmlObs}[1]{\langle#1\rangle}
\newcommand{\hmlEps}{\hmlObs{\varepsilon}}
\newcommand{\hmlOpt}[1]{\text{\rm\smaller(}#1\text{\rm\smaller)}}
\newcommand{\hmlTauOpt}{\hmlOpt{\tau}}
\newcommand{\hmlObsI}[1]{\hmlObs{\ccsInm{#1}}}
\newcommand{\hmlAnd}[2]{{\bigwedge_{#1 \in #2}}}
\newcommand{\hmlAndS}{{\bigwedge}}
\newcommand{\hmlTrue}{\mathsf{T}}
\newcommand{\hmlNeg}{\neg}
\newcommand{\hmlSemantics}[3]{{\llbracket #1 \rrbracket^{#2}_{#3}}}
\newcommand{\hmlStrategies}{\mathsf{Strat}}
\newcommand{\hmlPrune}{\mathsf{prune\_dominated}}
\newcommand{\height}{\mathsf{height}}
\newcommand{\expr}{\operatorname{\mathsf{expr}}}
\newcommand{\prices}{\mathbf{Pr}}

\let\obs\undefined
\newcommand*{\eqName}[1]{\mathrm{#1}}
\newcommand*{\obs}[1]{\mathcal{O}_{\mathrm{#1}}}
\newcommand*{\bEquiv}[1]{\sim_{\mathrm{#1}}}
\newcommand*{\bPreord}[1]{\preceq_{\mathrm{#1}}}

\newcommand{\refDef}[1]{Definition~\ref{#1}}
\newcommand{\refExample}[1]{Example~\ref{#1}}
\newcommand{\refRem}[1]{Remark~\ref{#1}}
\newcommand{\refThm}[1]{Theorem~\ref{#1}}
\newcommand{\refProp}[1]{Proposition~\ref{#1}}
\newcommand{\refLem}[1]{Lemma~\ref{#1}}
\newcommand{\refCor}[1]{Corollary~\ref{#1}}
\newcommand{\refSec}[1]{Section~\ref{#1}}
\newcommand{\refSubsec}[1]{Subsection~\ref{#1}}
\newcommand{\refFig}[1]{Figure~\ref{#1}}
\newcommand{\refClaim}[1]{Claim~\eqref{#1}}
\newcommand{\refAlgo}[1]{Algorithm~\ref{#1}}
\newcommand{\refLine}[1]{Line~\ref{#1}}

\newcommand*{\ie}{\text{i.e.\,}}
\newcommand*{\cf}{\text{cf.\,}}

\makeatletter
\newbox\xrat@below
\newbox\xrat@above
\newcommand{\xrightarrowtail}[2][]{%
  \setbox\xrat@below=\hbox{\ensuremath{\scriptstyle #1}}%
  \setbox\xrat@above=\hbox{\ensuremath{\scriptstyle #2}}%
  \pgfmathsetlengthmacro{\xrat@len}{max(\wd\xrat@below,\wd\xrat@above)+.6em}%
  \mathrel{\tikz [>->,baseline=-.58ex,line width=0.43pt]
                 \draw (0,0) -- node[below=-2pt] {\box\xrat@below}
                                node[above=-2pt] {\box\xrat@above}
                       (\xrat@len,0) ;}}
\makeatother

\newcommand{\subf}[2]{%
{\small\begin{tabular}[t]{@{}c@{}}
           #1\\
           #2
\end{tabular}}%
}